\documentclass[sigconf]{acmart}
\usepackage{tikz}

\usepackage{booktabs}
\usepackage{array}
\usepackage{balance} 
\usepackage{lipsum}
\usepackage{multirow}
\usepackage{booktabs}
\usepackage{amsmath}
\newcommand{\ie}{\emph{i.e., }}
\newcommand{\eg}{\emph{e.g., }}
\newcommand{\etal}{\emph{et al. }}

\newcommand{\wrt}{\emph{w.r.t. }}
\newcommand{\cf}{\emph{cf. }}

\usepackage{subfigure}

\usepackage{algorithm}  
\usepackage{algorithmicx}  
\usepackage{algpseudocode}  
\usepackage{amsmath}  
\usepackage{enumitem}

\begin{document}



\copyrightyear{2021}
\acmYear{2021}
\setcopyright{acmcopyright}\acmConference[SIGIR '21]{Proceedings of the 44th
International ACM SIGIR Conference on Research and Development in Information
Retrieval}{July 11--15, 2021}{Virtual Event, Canada}
\acmBooktitle{Proceedings of the 44th International ACM SIGIR Conference on
Research and Development in Information Retrieval (SIGIR '21), July 11--15, 2021,
Virtual Event, Canada}
\acmPrice{15.00}
\acmDOI{10.1145/3404835.3462962}
\acmISBN{978-1-4503-8037-9/21/07}

\fancyhead{}
\title{Clicks can be Cheating: Counterfactual Recommendation \\for Mitigating Clickbait Issue}

\author{Wenjie Wang}
\email{wenjiewang96@gmail.com}
\affiliation{%
  \institution{National University of Singapore}
  \country{}
}
\author{Fuli Feng}
\authornote{Corresponding author: Fuli Feng (fulifeng93@gmail.com).}
\email{fulifeng93@gmail.com}
\affiliation{%
  \institution{Sea-NExT Joint Lab, Singapore National University of Singapore}
  \country{}
}

\author{Xiangnan He}
\email{hexn@ustc.edu.cn}
\affiliation{%
  \institution{University of Science and Technology of China} \country{}
}

\author{Hanwang Zhang}
\email{hanwangzhang@ntu.edu.sg}
\affiliation{%
  \institution{Nanyang Technological University}
  \country{}
}

\author{Tat-Seng Chua}
\email{dcscts@nus.edu.sg}
\affiliation{%
  \institution{National University of Singapore}
  \country{}
}

\begin{abstract}

Recommendation is a prevalent and critical service in information systems. To provide personalized suggestions to users, industry players embrace machine learning, more specifically, building predictive models based on the click behavior data. This is known as the Click-Through Rate (CTR) prediction, which has become the gold standard for building personalized recommendation service. However, we argue that there is a significant gap between clicks and user satisfaction --- it is common that a user is ``cheated'' to click an item by the attractive title/cover of the item. This will severely hurt user's trust on the system if the user finds the actual content of the clicked item disappointing. What's even worse, optimizing CTR models on such flawed data will result in the Matthew Effect, making the seemingly attractive but actually low-quality items be more frequently recommended. 

In this paper, we formulate the recommendation models as a causal graph that reflects the cause-effect factors in recommendation, and address the clickbait issue by performing counterfactual inference on the causal graph. We imagine a counterfactual world where each item has only exposure features (\ie the features that the user can see before making a click decision). By estimating the click likelihood of a user in the counterfactual world, we are able to reduce the direct effect of exposure features and eliminate the clickbait issue. Experiments on real-world datasets demonstrate that our method significantly improves the post-click satisfaction of CTR models. 

\end{abstract}

%



\maketitle

\section{Introduction}

Recommender systems have been increasingly used to alleviate information overloading for users in a wide spectrum of information systems such as e-commerce~\cite{Zhang2020Large}, digital streaming~\cite{wei2019mmgcn}, and social networks~\cite{He2017Neural}. 
To date, the most recognized way for training recommender model is to optimize the Click-Through Rate (CTR), which aims to maximize the likelihood that a user clicks the recommended items. 
Despite the wide deployment of CTR optimization in recommender systems, we argue that the user experience may be hurt unintentionally due to the clickbait issue. 
That is, some items with attractive exposure features (\eg title and cover image) are easy to attract user clicks \cite{Hofmann2012on, Yue2010Beyond}, and thus are more likely to be recommended, but their actual content does not match the exposure features and disappoints the users. Such clickbait issue is very common, especially in the present era of self-media, posing great obstacles for the platform to provide high-quality recommendations (\cf Figure \ref{fig:data_explore} for the evidence).  

To illustrate, Figure \ref{example} shows an example that a user clicks two recommended videos with observation of their exposure features only. After watching the video, \ie examining the video content after clicking, the user gives the ratings of whether like or dislike the recommendations. $Item 2$ receives a dislike since the title deliberately misleads the user to click it, whereas $item 1$ receives a like since its actual content matches the title and cover image, and satisfies the user. 
This reflects the possible (in fact, significant) gap between clicks and satisfaction --- many clicks would end up with dissatisfaction since the click depends largely on whether the user is interested in the exposure features of the item.

Assuming that we can extract good content features that are indicative of item quality and even consistent with user satisfaction, can we address the discrepancy issue? Unfortunately, the answer is no. The reason roots in the optimization objective --- CTR: when we train a recommender model to maximize the click likelihood of the items with the clickbait issue, the model will learn to emphasize the exposure features and ignore the signal from other features, because the attractive exposure features are the causal reason of user clicks. This will aggravate the negative effect of clickbait issue, making these seemingly attractive but low-quality items be recommended more and more frequently.

To address the issue, a straightforward solution is to leverage the post-click feedback from users~\cite{Wen2019Leveraging, Lu2018Between}, such as the like/dislike ratings and numeric reviews. However, the amount of such explicit feedback is much smaller than that of click data, since many users are reluctant to leave any feedback after clicks. In most real-world datasets, users have very few post-click feedback, making it difficult to utilize them to supplement the large-scale implicit feedback well. Towards a wider range of applications and broader impact, we believe that it is critical to solve the clickbait issue in recommender system based on the click feedback only, which is highly challenging and has never been studied before.

In this work, we approach the problem from a novel perspective of causal inference: if we can distinguish the effects of exposure features (pre-click) and content features (post-click) on the prediction, then we can reduce the effect of exposure features that cause the clickbait issue. Towards this end, we first build a causal graph that reflects the cause-effect factors in recommendation scoring (Figure \ref{fig:causal_our}(b)). Next, we estimate the \textit{direct effect} of exposure features on the prediction score in a counterfactual world (Figure \ref{fig:causal_our}(c)), which imagines \textit{what the prediction score \textbf{would} be if the item \textbf{had} only the exposure features}. During inference, we remove this direct effect from the prediction in the factual world, which presents the total effect of all item features. In the example of Figure \ref{example}, although $item 1$ and $item 2$ obtain similar scores in the factual world, the final score of $item 2$ will be largely suppressed, because its content features are disappointing and it is the deceptive exposure features that increase the prediction score in the factual world. We instantiate the framework on MMGCN~\cite{wei2019mmgcn}, a representative multi-modal recommender model that can handle both exposure and content features. Extensive experiments on two widely used benchmarks show the superiority of the proposed framework, which significantly reduces the clickbait issue by only using the click feedback and recommends more satisfying items. 


To sum up, the contributions of this work are threefold:

\begin{itemize}[leftmargin=*]
    \item We highlight the importance of mitigating the clickbait issue by using click data only and leverage a new causal graph to formulate the recommendation process. 
    \item We introduce counterfactual inference into recommendation to mitigate the clickbait issue, and propose a counterfactual recommendation framework which can be applied to any recommender models with item features as inputs.
    \item We implement the proposed framework on MMGCN and conduct extensive experiments on two widely used benchmarks, which validate the effectiveness of our proposal.
\end{itemize}


\section{Task Formulation}
\label{sec:task}
In this section, we formulate the recommender training and the clickbait issue, followed by the task evaluation.

\vspace{3pt}
\textbf{Recommender training.}
The target of recommender training is to learn a scoring function $s_{\theta}$ that predicts the preference of a user over an item. 
Formally,
$Y_{u,i} = s_{\theta}(u, i)$
where $u$ and $i$ denote user features and item features, respectively. Specifically, item features $i = (e, t)$ include both exposure features $e$ and content features $t$ which are observed by users before and after clicks, respectively. $\theta$ denotes the model parameters which are typically learned from historical click data $\bar{\mathcal{D}} = \{(u, i, \bar{Y}_{u,i}) | u \in \mathcal{U}, i \in \mathcal{I}\}$, where $\bar{Y}_{u,i} \in \{0, 1\}$ denotes whether $u$ clicks $i$ ($\bar{Y}_{u,i}=1$) or not ($\bar{Y}_{u,i}=0$). $\mathcal{U}$ and $\mathcal{I}$ refer to the user set and item set, respectively. In this work, we use click to represent any type of implicit interactions for brevity, including purchase, watch, and download. 
Formally, the recommender training is:
\begin{equation}
\begin{aligned}
& \bar{\theta} =  \mathop{\arg\min}_{\theta}\sum_{(u, i,  \bar{Y}_{u,i}) \in \bar{\mathcal{D}}}l(s_\theta(u, i), \bar{Y}_{u,i}),
\end{aligned}
\label{eq:usual_obj}
\end{equation}
where $l(\cdot)$ denotes the recommendation loss such as cross-entropy loss \cite{Goodfellow2016deep}.
During inference, the trained recommender model serves each user by ranking all items according to $Y_{u, i} = s_{\bar{\theta}}(u, i)$ and recommending the top-ranked ones to the user.

\begin{figure}[tp]
\vspace{0cm}
\setlength{\abovecaptionskip}{0.2cm}
\setlength{\belowcaptionskip}{-0.2cm}
\includegraphics[scale=0.35]{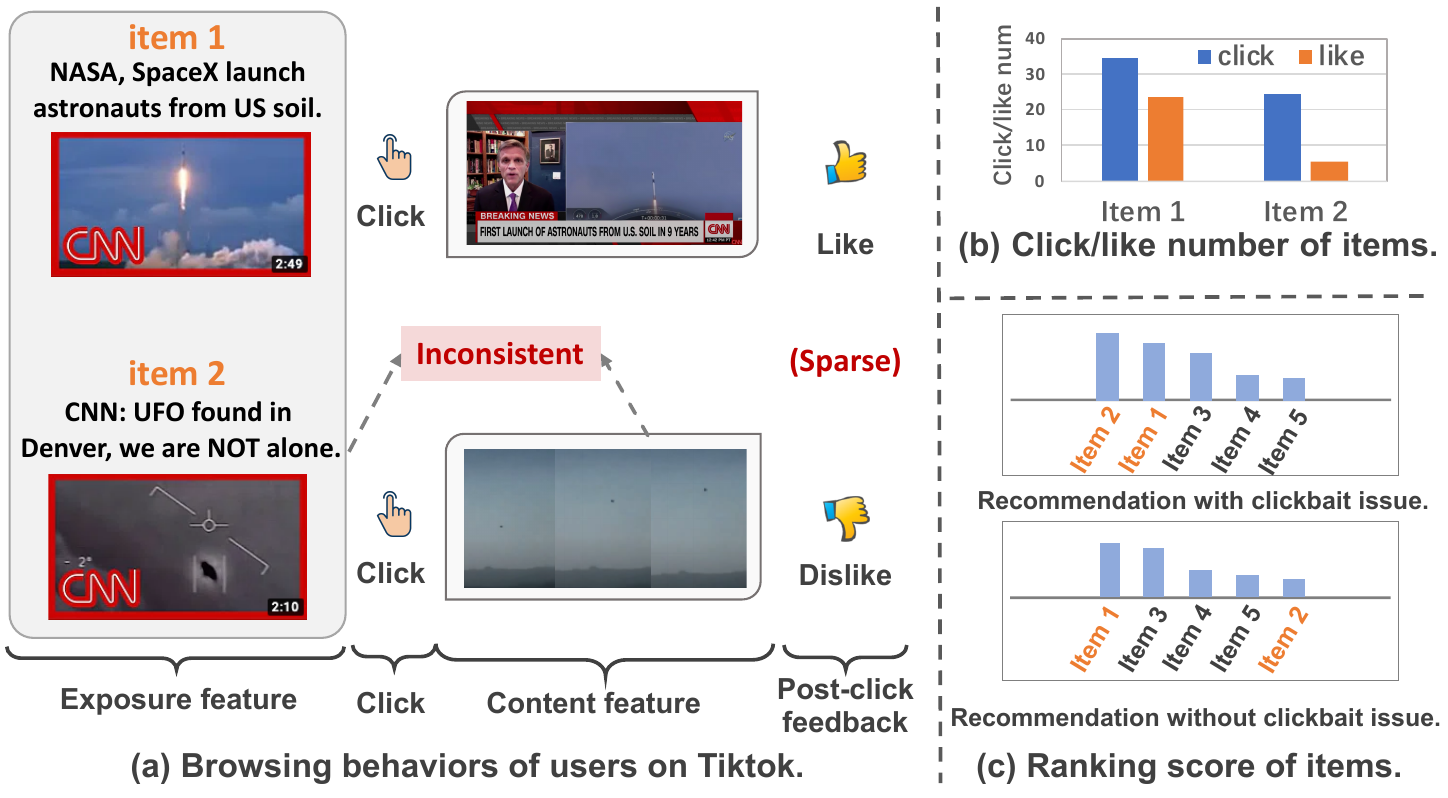}
\caption{(a) Illustration of inconsistency between clicks and likes. (b) Number of clicks/likes on the two items where few clicks on item 2 end with likes. (c) Two recommendation lists with and without the clickbait issue, respectively.}
\label{example}
\end{figure}


\vspace{3pt}
\textbf{Clickbait Issue.}
The clickbait issue is recommending items with attractive exposure features but disappointing content features frequently. 
Formally, given item $i$ with attractive exposure features but dissatisfying content, and item $j$ with less attractive exposure features and satisfying content, the clickbait issue happens if:
\begin{equation}
\begin{aligned}
& s_{\bar{\theta}}(u, i=(e_i, t_i)) > s_{\bar{\theta}}(u, j= (e_j, t_j)),
\end{aligned}
\label{eq:clickbait}
\end{equation}
where item $i$ ranks higher than item $j$.
That is, items with more attractive exposure features (\eg $item 2$ in Figure \ref{example}) occupy the recommendation opportunities of items with satisfying content features (\eg $item 1$ in Figure \ref{example}). 

Consequently, the recommender models will recommend many items like $i$, which will hurt user experience and lead to more clicks that end with dislikes. And worse still, it forms a vicious spiral: in turn, such clicks aggravate the issue in future recommender training.
In this work, we aim to break the vicious spiral by mitigating the clickbait issue during inference, \ie forcing $Y_{u, i} < Y_{u, j}$ for more user satisfaction rather than a higher CTR. Furthermore, we solve the problem based on click feedback only, \ie no post-click feedback is accessible during the recommender training. 

\vspace{3pt}
\textbf{Evaluation.}
Distinct from the conventional recommender evaluation that treats all clicks in the testing period as positive samples~\cite{wei2019mmgcn, he2016vbpr}, we evaluate recommendation performance only over clicks that end with positive post-click feedback (\ie likes)~\cite{wen2020entire}.
We do not use the clicks that lack post-click feedback due to the unawareness of user satisfaction. In addition, we believe that the recommendation performance on the selected clicks is able to validate the effectiveness of solving the clickbait issue. This is because a recommender model affected by the clickbait issue will fail on a portion of the selected clicks because they prefer to recommend items with more attractive exposure features but dissatisfying content features.

\begin{figure}[t]
\centering
\setlength{\abovecaptionskip}{0.1cm}
\setlength{\belowcaptionskip}{-0.3cm}
\includegraphics[scale=0.45]{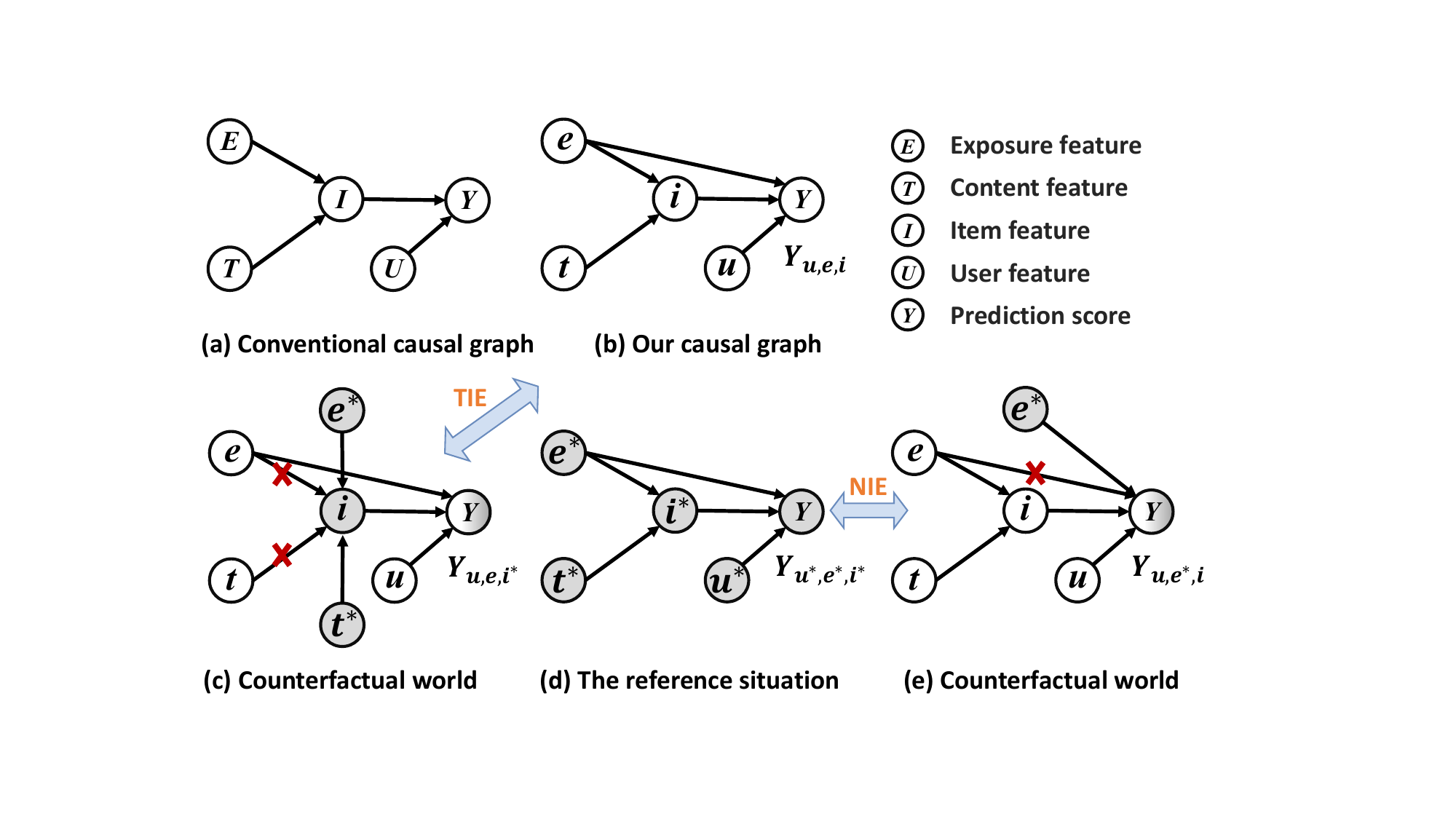}
\caption{(a) An example of a causal graph where the individual income ($I$) is directly affected by the education ($E$), age ($A$), and skill ($S$); and skill is influenced by education. (b) A causal graph with particular realizations. (c) A causal intervention $do(E=e^*)$, where $e^*$ denotes the reference value of $E$, \eg no qualifications. (d) One counterfactual where $S$ is set as $s^*$ while keeping $E=e$ on the edge $E \rightarrow I$.}
\label{fig:causal_graph}
\vspace{0cm}
\end{figure}

\section{Preliminary}
\label{sec:preliminary}
We briefly introduce the concepts of counterfactual inference~\cite{judea2001direct, pearl2009causality} used in this paper, and refer readers to learn from the related works \cite{vanderweele2013three, judea2001direct, tang2020unbiased, niu2020counterfactual, tang2020longtailed} for a comprehensive understanding.


\vspace{3pt}
\textbf{Causal Graph.}
Causal graph describes the causal relations between variables by a directed acyclic graph $\mathcal{G = \{N, E \}}$, where $\mathcal{N}$ is the set of variables (\ie nodes) and $\mathcal{E}$ records the causal relations (\ie edges). In the causal graph, capital letters and lowercase letters denote random variables (\eg $X$) and the specific realizations of random variables (\eg $x$), respectively. 
Figure~\ref{fig:causal_graph}(a) illustrates an example of a causal graph that represents the causal relations to the individual income: 1) the individual income ($I$) is directly affected by the education ($E$), age ($A$), and skill ($S$); and 2) indirectly affected by the education through a \textit{mediator} $S$. According to the graph structure, a set of structural equations $\mathcal{F}$~\cite{pearl2009causality} can be used to measure how the variables are affected by their parents. For example, we can estimate the values of $S$ and $I$ from their parents by $\mathcal{F}=\{f_S(\cdot), f_I(\cdot)\}$. Formally,
\begin{equation}
\left\{
\begin{aligned}
& S_e = s = f_S(E=e), \\
& I_{e,s,a}=f_I(E=e, S=s, A=a), 
\end{aligned}
\right .
\end{equation}
where $I_{e,s,a}$ denotes the income of one person who satisfies $E=e$, $S=s$, and $A=a$. $f_S(\cdot)$ and $f_I(\cdot)$ correspond to the structural equations of variable $S$ and $I$, respectively, which can be learned from a set of observations~\cite{pearl2009causality}. 


\vspace{3pt}
\textbf{Counterfactuals.}
Counterfactual inference \cite{Pearl2018the} is a technique to estimate what the descendant variables would be if the value of one \textit{treatment variable} were different with its real value in the factual world. 
As shown in Figure \ref{fig:causal_graph}(d), counterfactual inference can estimate \textit{what the income of Joe {would} be if he {only had} the skill of a person without qualifications}. That is imagining a situation: $I$ receives $E=e$ through $E\rightarrow I$, while $S$ receives $E=e^*$ through $E\rightarrow S$ and other variables are fixed. Specifically, $e$ can represent a bachelor degree while $e^*$ denotes no qualifications.
The key to counterfactual inference lies in performing external intervention~\cite{pearl2009causality} to control the value of $S$, which is termed as \textit{do-operator}. Formally, $do(S=s^*)$ forcibly substitute $s$ with $s^*=f_S(E=e^*)$ in the structural equation $f_I$, obtaining $I_{e,s^*,a} = f_I(E=e, S=s^*, A=a)$.
Note that $do(S=s^*)$ does not affect the ascendant variables of $S$, \ie $E$ retains its real value $e$ on the direct path $E\rightarrow I$.

\vspace{3pt}
\textbf{Causal Effect.}
Causal effect\footnote{In this work, causal effect is defined at the unit level~\cite{pearl2009causality, judea2001direct}, \ie the effect is on one individual rather than a population.} of one event with the treatment variable (\eg $E=e$, obtaining a bachelor degree) on the response variable (\eg $I$) measures the change of the response variable when the treatment variable changes from its reference value (\eg $e^*$) to the expected value (\eg $e$), which is also termed as \textit{total effect} (TE).
Formally, the TE of $E=e$ on $I$ under situation $A=a$ is defined as:
\begin{equation}
\begin{aligned}
    \text{TE} & = I_{e, s, a} - I_{e^*, s^*, a}, \\
    & = f_I(E=e, S=s, A=a) - f_I(E=e^*, S=s^*, A=a),
\end{aligned}
\end{equation}
where $I_{e^*, s^*, a}$ denotes the reference status of $I$ when $E=e^*$, \ie the outcome of the intervention $do(E=e^*)$ (see Figure \ref{fig:causal_graph}(c)). Specifically, by viewing $e^*$ as no qualifications, $I_{e^*, s^*, a}$ denotes the income of Joe if he hadn't got qualifications (\ie $E=e^*$) at the age of $a$.
%
Furthermore, the event affects the response variable through both the direct path between the two variables (\eg $E \rightarrow I$) and the indirect path via mediators (\eg $E \rightarrow S \rightarrow I$). A widely used decomposition of TE is TE = NDE + TIE, where NDE and TIE denote the \textit{natural direct effect} and \textit{total indirect effect}~\cite{judea2001direct, vanderweele2013three}, respectively. 

In particular, NDE is the change of the response variable when only changing the treatment variable on the direct path, \ie the mediators retain unchanged and still receive the reference value. For instance, the NDE of $E=e$ on $I$ under situation $a$ is the change of the income $I$ when changing $E$ from $e^*$ to $e$ and forcing $S=s^*$. Formally, the calculation of NDE relies on $do(S=s^*)$, which is:
\begin{equation}
\begin{aligned}
\text{NDE} = I_{e, s^*, a} - I_{e^*, s^*, a},
\end{aligned}
\label{eq:nde_pre}
\end{equation}
where $I_{e, s^*, a}$ is the income in a counterfactual world (see Figure \ref{fig:causal_graph}(d)). Accordingly, the TIE of $E=e$ on $I$ under situation $A=a$ can be obtained by subtracting NDE from TE~\cite{vanderweele2013three}:
\begin{equation}
\begin{aligned}
\text{TIE} = \text{TE} - \text{NDE} = I_{e, s, a} - I_{e, s^*, a}.
\end{aligned}
\label{eq:tie_pre}
\end{equation}
Generally, TIE is the change of the response variable when the mediators are changed from their reference values (\eg $s^* = f_S(E=e^*)$) to the ones receiving the expected value (\eg $s=f_S(E=e)$), and the value of the treatment variable on the direct path remains fixed (\eg $E=e$ on $E\rightarrow I$).

\section{Counterfactual Recommendation}\label{sec:method}
In this section, we introduce the causal graph of recommender systems, followed by the elaboration of counterfactual inference to mitigate the clickbait issue and the design of proposed counterfactual recommendation (CR) framework.

\subsection{Causal Graph of Recommender Systems}
\begin{figure}[tb]
\setlength{\abovecaptionskip}{0.1cm}
\setlength{\belowcaptionskip}{-0.5cm}
\centering
\includegraphics[scale=0.45]{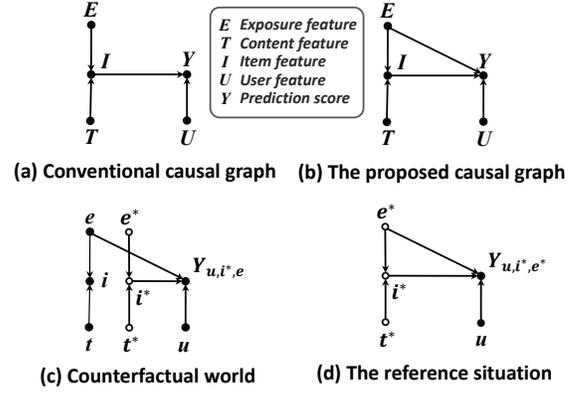}
\caption{The causal graphs for conventional and counterfactual recommendations. $*$ denotes the reference values.}
\label{fig:causal_our}
\end{figure}

In Figure~\ref{fig:causal_our}(a), we abstract the causal graph of existing recommender models where $Y$, $U$, $I$, $E$, and $T$ denote the prediction score, user features, item features, exposure features, and content features, respectively. Accordingly, the existing recommender model (\ie $s_{\theta}(\cdot)$) is abstracted as two structural equations $f_Y(\cdot)$ and $f_I(\cdot)$, which are formulated as:
\begin{equation}
\begin{aligned}
Y_{u, i} = f_Y(U=u, I=i),~where~i = I_{e,t} = f_I(E=e, T=t).
\end{aligned}
\end{equation}
The two structural equations $f_Y(\cdot)$ and $f_I(\cdot)$ correspond to the main modules of the existing models, the scoring function (\eg inner product function) and feature aggregation function (\eg multi-layer perceptron (MLP)~\cite{Goodfellow2016deep}), respectively. In particular, $f_I(\cdot)$ aims to extract the representative item features from its exposure and content features, which are then fed into $f_Y(\cdot)$ for making the prediction. 
The parameters of the equations (\ie $\theta$) are learned by minimizing the recommendation loss over historical data, so as to maximize the likelihood of the clicked items (\ie Equation~\ref{eq:usual_obj}). 

However, the causal graph of existing recommender models mismatches the generation process of the training data. 
In the user browsing process, users might click the items only because they are attracted by the exposure features\footnote{Note that the click behavior can also be affected by other item features (\ie $I$), \eg the category and uploader of videos.}. From the cause-effect view, there is a direct effect from the exposure features to the click behavior.
As a result of ignoring such direct effect in the model, the feature aggregation function will inevitably emphasize the exposure features while ignoring the content features (see empirical results in Figure~\ref{fig:synthetic}), in order to achieve a small loss on the clicked items with the clickbait issue.


To bridge this gap, we build a new causal graph by adding a direct edge from exposure features $E$ to the prediction $Y$ (Figure \ref{fig:causal_our}(b)).
%
According to the new causal graph, the recommender model should capture the causal effect of exposure features on prediction $Y$ through both the direct path ({$E$} $\rightarrow$ {$Y$}) and the indirect path ({$E$} $\rightarrow$ {$I$} $\rightarrow$ {$Y$}).
Formally, the abstract format of the model should be:
\begin{equation}
\begin{aligned}
Y_{u, i, e} = f_Y(U=u, I=i, E=e),~where~i = f_I(E=e, T=t).
\end{aligned}
\label{eq:structuraleqs}
\end{equation}
In other words, when we design a recommender model that will be optimized over historical clicks through the CTR objective, its scoring function should directly take exposure features as one additional input. 

\subsection{Mitigating Clickbait Issue}
While the new causal graph provides a more precise description of the cause-effect factors for recommendation scoring, the recommender model based on the new causal graph still suffers from the clickbait issue (in Equation~\ref{eq:clickbait}). This is because the outcome of the response variable, \ie $Y_{u, i, e}$, still accounts for the direct effect of exposure features. Consequently, the item (\eg \textit{item2} in Figure~1) with more attractive exposure features is still scored higher than the one with more satisfying content but less attractive exposure features. To mitigate the clickbait issue, we perform \textit{CR inference} to reduce the direct effect of exposure features from the prediction $Y_{u, i, e}$, which is formulated as $Y_{u, i, e} - \text{NDE}$.

\vspace{1pt}
Towards this end, we need to estimate the NDE of event $E=e$ on the response variable $Y$. In particular, we estimate the NDE under situation $U=u$ and $T=t^*$. 
As detailed in Section \ref{sec:preliminary}, 
the NDE is formulated as:
\begin{equation}\notag
\begin{aligned}
\text{NDE} &= Y_{u, i^*, e} - Y_{u, i^*, e^*} \\
&= f_Y(U=u,I=i^*,E=e) - f_Y(U=u,I=i^*,E=e^*),
\end{aligned}
\end{equation}
where $i^*=f_I(E=e^*, T=t^*)$, and $e^*$ and $t^*$ are the reference values of $E$ and $T$, respectively. $f_Y(U=u,I=i^*,E=e)$ denotes the outcome of a counterfactual (see Figure \ref{fig:causal_our}(c)) where the treatment variable $E$ is changed from $e^*$ to $e$ on the direct path (\ie $E\rightarrow Y$) while remains its reference value on the indirect path (\ie $E\rightarrow I\rightarrow Y$).
That is, it estimates \textit{what the prediction score would be if the item had only the exposure features} in a counterfactual world, \ie to what extent the user is purely attracted by exposure features. 
In this task, the reference values $e^*$ and $t^*$ are treated as the status that the features are not given. 
Given the user features $u$, the second term $Y_{u, i^*, e^*}$ (Figure \ref{fig:causal_our}(d)) is thus a constant for any items, \ie $Y_{u, i^*, e^*}$ will not affect the ranking of items for a user.
%
%
Therefore, by subtracting the NDE of exposure features from $Y_{u, i, e}$, the prediction score of CR inference becomes:
\begin{equation}
\begin{aligned}
Y_{CR} =  Y_{u, i, e} - Y_{u, i^*, e}.
\end{aligned}
\end{equation}

Intuitively, $Y_{CR}$ reduces the NDE of exposure features and relies on the effect of the combined item features $I$ for inference. 
The prediction score of the item with attractive exposure features but boring content (\eg $item 2$ in Figure \ref{example}) will be largely suppressed during CR inference,
because its only attractiveness is in the exposure features and the content features are dissatisfying. It will have a high prediction score in the counterfactual world (\ie $Y_{u, i^*, e}$).
Accordingly, the item with less attractive exposure features but satisfying content features (\eg $item 1$ in Figure \ref{example}) will have a higher chance to be recommended because the satisfactory item features $I$ will increase the prediction score in CR inference, which forces $s_{\theta}(u,i)<s_{\theta}(u,j)$ in Equation \ref{eq:clickbait}.

\vspace{1pt}
From the cause-effect view, CR inference subtracts the NDE of $E=e$ from the TE of $E=e$ and $T=t$. As introduced in Section \ref{sec:preliminary}, the TE of $E=e$ and $T=t$ on $Y$ under situation $U=u$ can be calculated by $Y_{u, i, e} - Y_{u, i^*, e^*}$ where $Y_{u, i^*, e^*}$ is the reference situation. Obviously, the prediction score of CR inference can be formulated as $Y_{CR} = \text{TE} - \text{NDE}$.

\vspace{1pt}
Note that we can estimate the NDE of $E=e$ on $Y$ under the situation of $T=t^*$ or $T=t$~\cite{judea2001direct}. Changes of the situation can lead to minor difference in the estimation since the recommender models are typically non-linear~\cite{vanderweele2013three, pearl2009causality}. We select the situation of $T=t^*$ to avoid the leakage of exposure features. 
This is because, in the recommendation scenarios, the content features $t$ might include some information in the exposure features $e$.
For instance, the cover image may be a frame in the video, which might cause the leakage of $e$ through the mediator $I$.
Empirical evidence in Table \ref{tab:effect_T} justifies the advantage of this choice.

\subsection{CR Framework Design}
Recall that the key to counterfactual inference lies in the learned structural equations. 
To enable CR inference, we thus need to design a recommender model according to the proposed causal graph in Figure \ref{fig:causal_our}(b) and an algorithm to learn the model parameters. 

\subsubsection{\textbf{Model Design}}
According to Equation~\ref{eq:structuraleqs}, the recommender model should consists of two functions: the scoring function $f_Y(U=u,I=i,E=e)$ and the feature aggregation function $f_I(E=e, T=t)$. As to the feature aggregation function, we can simply employ the one in existing models to encode the causal relations from $E$ and $T$ to $I$. 
We focus on upgrading the conventional scoring function $f_Y(U=u, I=i)$ to $f_Y(U=u,I=i,E=e)$.

\vspace{3pt}
$\bullet$ \textbf{Scoring Function.}
A straightforward idea is to embed the additional input $e$ into the conventional scoring function. However, this solution loses generality due to requiring careful adjustments for different recommender models.
According to the universal approximation theorem~\cite{csaji2001approximation}, we could also implement $f_Y(\cdot)$ by a MLP with $u$, $i$, and $e$ as the inputs. Nevertheless, it is hard to tune a MLP to achieve the comparable performance with the models wisely designed for the recommendation task~\cite{rendle2009bpr, He2017Neural, wei2019mmgcn}. 

Aiming to keep generality and leverage the advantages of existing models, the scoring function is implemented in a late-fusion manner~\cite{cadene2019rubi, niu2020counterfactual}:
\begin{equation}\notag
\begin{aligned}
f_Y(U=u,I=i,E=e) = f(Y_{u, i}, Y_{u, e}),
\end{aligned}
\end{equation}
where $Y_{u, i} = f_Y(U=u, I=i)$ and $Y_{u, e} = f_Y(U=u, E=e)$ are the predictions from two conventional models with different inputs; and $f(\cdot)$ is a fusion function. 
$Y_{u,i}$ and $Y_{u,e}$ can be instantiated by any recommender models with user and item features as the inputs such as MMGCN \cite{wei2019mmgcn} and VBPR \cite{he2016vbpr}.
In this way, we can simply adapt an existing recommender model to fit in the proposed causal graph by additionally implementing a fusion strategy, which can be easily achieved.

\vspace{3pt}
$\bullet$ \textbf{Fusion strategy.}
Inspired by the prior studies \cite{cadene2019rubi, niu2020counterfactual}, we adopt one classic fusion strategy: Multiplication (MUL), formulated as: 
\begin{equation}\notag
\begin{aligned}
&Y_{u, i, e} = f_Y(U=u, I=i, E=e) = f(Y_{u, i}, Y_{u, e}) = Y_{u, i} * \sigma(Y_{u, e}),
\end{aligned}
\end{equation}
where $\sigma$ denotes the sigmoid function.
It provides non-linearity for sufficient representation capacity of the fusion strategy, which is essential (see results in Table \ref{tab:FusionStrategy}).
Note that the proposed CR is general to any differentiable arithmetic binary operations and we compare more strategies in Table \ref{tab:FusionStrategy}.

\subsubsection{\textbf{Model Training}}
Recall that the CR inference requires two predictions: $Y_{u, i, e}$ and $Y_{u, i^*, e}$. The target of model training is thus twofold --- learning parameters of the structural equations (\ie $f_Y(\cdot)$ and $f_I(\cdot)$) that can accurately estimate both $Y_{u, i, e}$ and $Y_{u, i^*, e}$. As such, we minimize a multi-task training objective over historical clicks to learn the model parameters, which is formulated as:
\begin{equation}\small
\begin{aligned}
& \sum_{(u, i,  \bar{Y}_{u,i}) \in \mathcal{\bar{D}}}l(Y_{u, i, e}, \bar{Y}_{u,i}) + \alpha * l(Y_{u, e}, \bar{Y}_{u,i}),
\end{aligned}
\end{equation}
where $\bar{Y}_{u,i}$ is the label for $u$ and $i$, and $\alpha$ is a hyperparameter to tune the relative weight of two tasks.
Recall that $i^*$ indicates the recommender model doesn't take $i$ as the input, and thus $Y_{u, e}$ can be seen as the learned prediction $Y_{u, i^*, e}$ based on the user features $u$ and exposure features $e$ in the counterfactual world.

\vspace{3pt}
$\bullet$ \textbf{CR Inference.}
CR inference needs to calculate the predictions $Y_{u, i, e}=f(Y_{u, i}, Y_{u, e})$ and $Y_{u, i^*, e} =f(c_{u}, Y_{u, e})$ where $c_{u}$ refers to the expectation constants of $Y_{u, {I}}$:
\begin{equation}
\begin{aligned}
& c_{u} = E(Y_{u, {I}}) = \frac{1}{|\mathcal{I}|} \sum_{i\in \mathcal{I}} Y_{u, i},
\end{aligned}
\end{equation}
which indicates that for each user, all the items share the same score $c_u$. Since the features of $I$ are not given in $Y_{u, i^*, e}$, the model used to predict $Y_{u,i}$ ranks items with the same score $c_u$ for user $u$. 
In this way, the results of CR inference will be calculated by:
\begin{equation}
\notag
\begin{aligned}
Y_{CR} &= Y_{u, i, e} - Y_{u, i^*, e} = Y_{u, i, e} - f(c_{u}, Y_{u, e})
= Y_{u, i, e} - c_{u} * \sigma(Y_{u, e}).
\end{aligned}
\end{equation}
The item with the attractive exposure features but dissatisfying content will have a higher score of $Y_{u, e}$, which is then subtracted from the original prediction $Y_{u, i, e}$, lowering the rank of such items.

To summarize, compared to conventional recommender models, the proposed CR framework demonstrates three main differences:
\begin{itemize}[leftmargin=*]
    \item \textbf{Causal graph.} The recommender model under the CR framework is based on a new causal graph that accounts for the direct effect of exposure features on the prediction score.
    \item \textbf{Multi-task training.} In addition to the model learning in the real world (\ie $Y_{u,i,e}$), we also train the model to make predictions in the counterfactual world (\ie $Y_{u,i^*,e}$).
    \item \textbf{CR inference.} Instead of making recommendations according to the real-world prediction, we deduct the NDE of exposure features to mitigate the clickbait issue.
\end{itemize}

\section{Related Work}


\vspace{3pt}
\noindent\textbf{Recommendation.}
Because of the rich user/item features in the real-world scenarios \cite{hong2017coherent, hong2015learning, liu2020iterative}, many approaches \cite{Chen2018temporal, li2019routing} incorporate multi-modal user and item features into recommendation \cite{he2016vbpr,jiang2020aspect, chen2017attentive, Wang2021Market2Dish}. 
Recently, Graph Neural Networks (GNN)~\cite{feng2019graph, feng2018learning} have been widely used in recommendation \cite{fan2019graph, wang2019NGCF, wei2020graph}, and GNN-based multi-modal model MMGCN \cite{wei2019mmgcn} achieves promising performance due to its modality-aware information propagation over the user-item graph. However, existing works are trained by implicit feedback and totally ignore the clickbait issue. Therefore, items with many clicks but few likes will be recommended frequently.

\vspace{3pt}
\noindent\textbf{Incorporating Various Feedback.}
To mitigate the clickbait issue, many efforts try to reduce the gap between clicks and likes by incorporating more features into recommendation, such as interaction context \cite{Kim2014Modeling}, item features \cite{Lu2019effects}, and various user feedback \cite{Yang2012Exploiting, yuan2020parameter}. Generally, they fall into two categories. The first is negative experience identification \cite{zannettou2018good, Lu2018Between}. It performs a two-stage pipeline \cite{Lu2018Between, Lu2019effects} which first identifies negative interactions based on item features (\eg the news quality) and context information (\eg dwell time), and then only uses interactions with likes as positive samples. 
The second category considers directly incorporating extra post-click feedback (\eg thumbs-up, favorite, and dwell time) to optimize recommender models~\cite{Yang2012Exploiting, liu2010understanding, Yi2014Beyond, Yin2013silence}. For instance, Wen \etal \cite{Wen2019Leveraging} leveraged the ``skip'' patterns to train recommender models with three kinds of items: ``click-complete'', ``click-skip'', and ``non-click''. 
Nevertheless, the application of these methods is limited by the availability of context information and users' additional post-click feedback. 
Post-click feedback is usually sparse,
and thus using only clicks with likes for training will lose a large proportion of positive samples. 

\vspace{3pt}
\noindent\textbf{Causal Recommendation.}
In the information retrieval domain, early studies \cite{ai2018unbiased, Thorsten2017unbiased} on causal inference mainly focus on de-biasing implicit feedback, \eg position bias \cite{Nick2008an}. As to causal recommendation \cite{Zhu2020unbiased, Konstantina2020Deconfounding, chen2020bias}, many researchers study fairness \cite{Marco2020Controlling} or the bias issues with the help of causal inference, such as exposure bias \cite{bonner2018causal,liang2016modeling} and popularity bias \cite{abdollahpouri2019managing} in the logged data \cite{Liu2020a}.
Among the family of causal inference for de-biasing recommendation, the most popular method is \textit{Inverse Propensity Scoring Weighting} (IPW) \cite{saito2020unbiased, ROSENBAUM1983the, liang2016causal} which turns the observed logged data into a pseudo-randomized trial by re-weighting samples. In general, they estimate the propensity of exposure or popularity at first, and re-weight samples with the inverse propensity scores. 
However, current causal recommendation never considers the clickbait issue. They don't distinguish the effects of exposure and content features, and treat users' implicit feedback such as clicks as the actual user preference. Therefore, prior studies still have the clickbait issue and recommend the items that many users would click but actually dislike.


\section{Experiments}
\label{sec:results}

\subsection{Experimental Settings}

\noindent\textbf{Datasets.}
We evaluate our proposed CR framework on two publicly available datasets in different application scenarios: Tiktok \cite{wei2019mmgcn} and Adressa \cite{Gulla2017the}. For each dataset, we utilize post-click feedback to evaluate the recommender models. We admit that: 1) the sparsity of post-click feedback might restrict the scale of the evaluation, however, we still cover a large group of users for evaluation. Actually, almost all users are covered in two datasets; and 2) the items with more attractive exposure features are easier to be collected as testing samples regardless of content features since they are more likely to be clicked. Nevertheless, constructing a totally unbiased testing set is unrealistic without external intervention, which is extremely expensive and thus left to future work. 
The statistics of datasets are in Table \ref{tab:statistic}. 

\begin{itemize}[leftmargin=*]
\item \textbf{Tiktok}. It is a multi-modal micro-video dataset released in ICME Challenge 2019\footnote{\url{http://ai-lab-challenge.bytedance.com/tce/vc/.}} where a micro-video has the features of caption, audio, and video. Multi-modal item features have already been extracted by the organizer for the fair comparison. We treat captions as exposure features and the remaining as content ones. Besides, actions of thumbs-up, favorite, or finish are used as the positive post-click feedback (\ie like), which is only used to construct the testing set for evaluation.

\item \textbf{Adressa}\footnote{\url{http://reclab.idi.ntnu.no/dataset/.}}. This is a news dataset~\cite{Gulla2017the} where the title and description of news are exposure features and the news content is treated as content features. We use the pre-trained Multilingual BERT \cite{devlin2018bert} to extract textual features into 768-dimension vectors. Following prior studies \cite{Kim2014Modeling}, we treat a click with dwell time > 30 seconds as a like of user.

\end{itemize}

\begin{table}[t]\small
\setlength{\abovecaptionskip}{0cm}
\setlength{\belowcaptionskip}{-0.1cm}
\centering
\caption{Statistics of two datasets.}\label{tab:statistic}
\begin{tabular}{l|c|c|c|c}
\toprule
\textbf{Dataset} & \textbf{\#Users} & \textbf{\#Items} & \textbf{\#Clicks} & \textbf{\#Likes} \\ \hline
\textbf{Tiktok}  & 18,855& 34,756& 1,493,532  & 589,008  \\ \hline
\textbf{Adressa} & 31,123& 4,895 & 1,437,540  & 998,612    \\ \bottomrule
\end{tabular}
\vspace{-0.3cm}
\end{table}

\begin{figure}
\vspace{0cm}
\setlength{\abovecaptionskip}{0cm}
\setlength{\belowcaptionskip}{-0.4cm}
\includegraphics[scale=0.22]{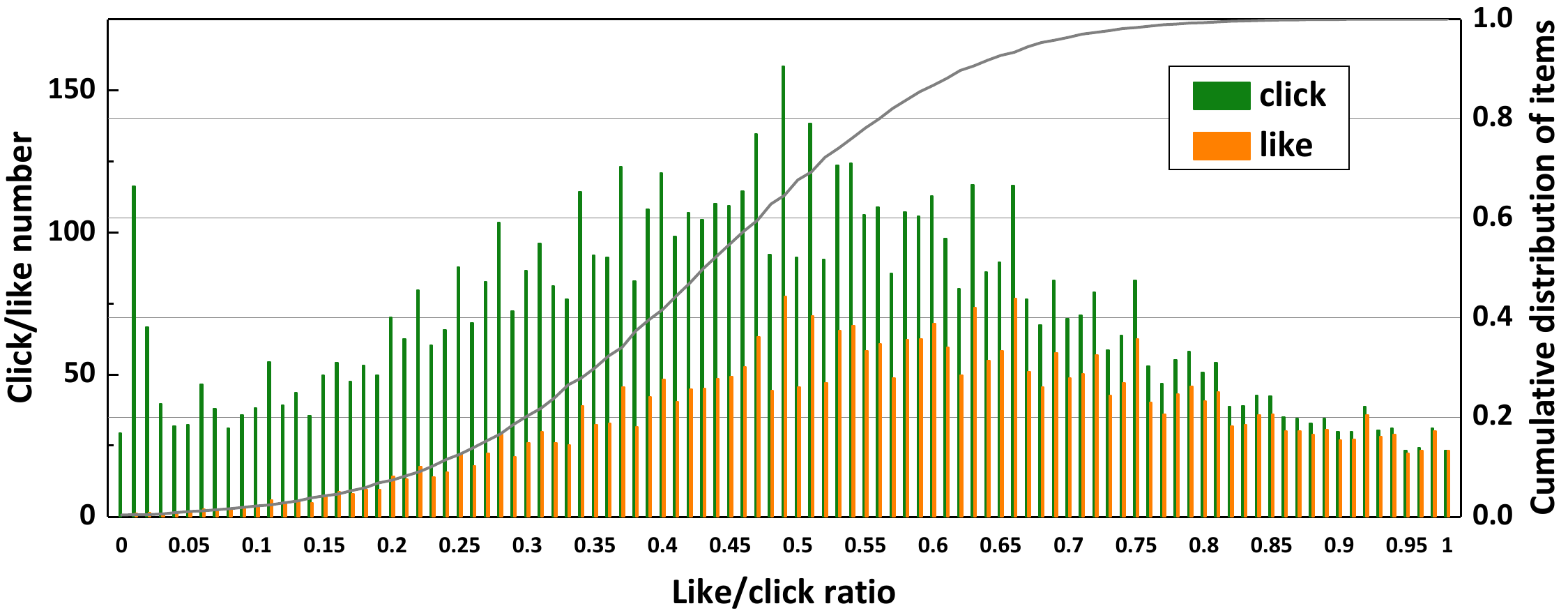}
\caption{Click and like distributions of items in Tiktok. The grey line visualizes the cumulative proportion of items as the like/click ratio increases. The x-axis is like/click ratio and the y-axis is the number of clicks or likes.}
\label{fig:data_explore}
\end{figure}

\begin{table*}[htbp]
\vspace{-0.1cm}
\setlength{\abovecaptionskip}{0cm}
\setlength{\belowcaptionskip}{0cm}
\caption{Top-$K$ recommendation performance of compared methods on Tiktok and Adressa. \%Improve. denotes the relative performance improvement of CR over NT. The best results are highlighted in bold. Stars and underlines denote the best results of the baselines with and without using additional post-click feedback during training, respectively.}
\label{tab:Results}
\begin{center}
\resizebox{\textwidth}{22mm}{
\begin{tabular}{l|lll|lll|lll|lll}
\hline 
\multicolumn{1}{c|}{\textbf{Dataset}} & \multicolumn{6}{c|}{\textbf{Tiktok}} & \multicolumn{6}{c}{\textbf{Adressa}} \\
\multicolumn{1}{c|}{\textbf{Metric}}  & \textbf{P@10} & \textbf{R@10} & \textbf{N@10} & \textbf{P@20} & \textbf{R@20} & \multicolumn{1}{l|}{\textbf{N@20}} & \textbf{P@10} & \textbf{R@10} & \textbf{N@10} & \textbf{P@20} & \textbf{R@20} & \textbf{N@20} \\ \hline \hline
\textbf{NT}~\cite{wei2019mmgcn}  & \underline{{0.0256}} & \underline{{0.0357}} & 0.0333& \underline{{0.0231}} & \underline{{0.0635}}  & 0.0430& \underline{{0.0501}} & \underline{{0.0975}}  & \underline{{0.0817}}& \underline{{0.0415}} & \underline{{0.1612}}  & \underline{{0.1059}}\\
\textbf{CFT}~\cite{wei2019mmgcn}  & 0.0253 & 0.0356  & \underline{{0.0339}}& 0.0226 & 0.0628  & \underline{{0.0437}}  & 0.0482 & 0.0942  & 0.0780 & 0.0405 & 0.1573  & 0.1021\\
\textbf{IPW}~\cite{liang2016causal}    & 0.0230  & 0.0334  & 0.0314& 0.0210  & 0.0582  & 0.0406  & 0.0419 & 0.0804  & 0.0663& 0.0361 & 0.1378  & 0.0883\\ \hline
\textbf{CT}~\cite{wei2019mmgcn}     & 0.0217 & 0.0295  & 0.0294& 0.0194 & 0.0520   & 0.0372  & 0.0493 & 0.0951  & 0.0799& $0.0418^{*}$ & 0.1611  & 0.1051\\
\textbf{NR}~\cite{Wen2019Leveraging}     & 0.0239 & 0.0346  & 0.0329& 0.0216 & 0.0605  & 0.0424  &  0.0499 & 0.0970  & 0.0814& 0.0415  & 0.1610  & 0.1058\\
\textbf{RR}    & $0.0264^{*}$ & $0.0383^{*}$  & $0.0367^{*}$ & $0.0231^{*}$ & $0.0635^{*}$  & $0.0430^{*}$  & $0.0521^{*}$ & $0.1007^{*}$  & $0.0831^{*}$ & 0.0415 & $0.1612^{*}$  & $0.1059^{*}$\\
\hline
\textbf{CR} & \textbf{0.0269} & \textbf{0.0393}  & \textbf{0.0370} & \textbf{0.0242} & \textbf{0.0683}  & \textbf{0.0476}  & \textbf{0.0532}   & \textbf{0.1045}    & \textbf{0.0878}  & \textbf{0.0439} & \textbf{0.1712}  & \textbf{0.1133}  \\
{\%Improve.} & 5.08\% & 10.08\% & 11.11\%  & 4.76\% & 7.56\%  & 10.70\%     & 6.19\% & 7.18\%  & 7.47\%& 5.78\% & 6.20\%  & 6.99\%  \\ 
\hline
\end{tabular}}
\end{center}
\end{table*}

Figure \ref{fig:data_explore} outlines the distribution of the like/click ratio where items are ranked and divided into 101 groups according to the ratio value. As can be seen, over 60\% of items have like/click ratio smaller than 0.5, indicating the wide existence of clicks that end with dislikes. Moreover, recommending such items may lead to more clicks which fail to satisfy users and hurt user experience.

For each user, we randomly choose 10\% clicks that end with likes to constitute a test set\footnote{If fewer than 10\% clicks of a user end with likes, all such clicks are put into the test set. Besides, we ignore the potential noise in the test set, \eg fake favorite.}, and treat the remaining as the training set. Besides, 10\% of clicks are randomly selected from the training set as the validation set. We utilize the validation set to tune hyper-parameters and choose the best model for the testing phase. For each click, we randomly choose an item the user has never interacted with as the negative sample for training.

\vspace{5pt}
\noindent\textbf{Evaluation Metrics.}
We follow the all-ranking evaluation protocol that ranks over all the items for each user except the clicked ones used in training~\cite{wang2021denoising, LightGCN2020he}, and report the recommendation performance through: Precision@K (P@K), Recall@K (R@K) and NDCG@K (N@K) with $K=\{10, 20\}$ where higher values indicate better performance \cite{wei2019mmgcn}.

\vspace{5pt}
\noindent\textbf{Compared Methods} 
We compare the proposed CR with various recommender methods that might alleviate the clickbait issue. For a fair comparison, all methods are applied to MMGCN \cite{wei2019mmgcn}, which is the state-of-the-art multi-modal recommender model and captures the modality-aware high-order user-item relationships. Specifically, CR is compared with the following baselines:

\begin{itemize}[leftmargin=*]

\item \textbf{NT.} Following \cite{wei2019mmgcn}, MMGCN is trained by the normal training (NT) strategy, where all item features are used and MMGCN is optimized with click data. We keep the same hyperparameter settings as in \cite{wei2019mmgcn}, including that: the model is optimized by the BPR loss~\cite{rendle2009bpr}; the learning rate is set as 0.001, and the size of latent features is 64.

\item \textbf{CFT.} Based on the analysis that exposure features are easy to induce the clickbait issue, we only use content features for training (CFT). The model is also trained with all click data.

\item \textbf{IPW.} Liang \emph{et al.}~\cite{liang2016causal, liang2016modeling} tried to reduce the exposure bias from clicks by causal inference with IPW~\cite{ROSENBAUM1983the}. For a fair comparison, we follow the idea of Liang \emph{et al.} and implement the exposure and click models in \cite{liang2016causal} by MMGCN since it uses multi-modal item features and thus can achieve better performance.
\end{itemize}

Besides, considering post-click feedback can indicate the actual user satisfaction, we compare CR with three baselines that additionally incorporate post-click feedback:

\begin{itemize}[leftmargin=*]
\item \textbf{CT.} This method is conducted in the clean training (CT) setting, in which only the clicks that end with likes are viewed as positive samples to train MMGCN.

\item \textbf{NR.} Wen \emph{et al.}~\cite{Wen2019Leveraging} adopted post-click feedback and also treated ``click-skip'' items as negative samples. We apply their Negative feedback Re-weighting (NR) into MMGCN. In detail, NR adjusts the weights of two negative samples during training, including ``click-skip'' items and ``no-click'' items. Following \cite{Wen2019Leveraging}, the extra hyper-parameter $\lambda_{p,n}$, \ie the ratio of two kinds of negative samples, is tuned in $\{0, 0.2, 0.4, 0.6, 0.8, 1.0\}$. 

\item \textbf{RR.} For each user, we propose a strategy to re-rank (RR) the top 20 items recommended by NT during inference. For each item, the final ranking is calculated by the sum of rank in NT and the rank based on the like/click ratio of items. The like/click ratio is calculated from the whole dataset.

\end{itemize}
We omit potential testing recommender models such as VBPR~\cite{he2016vbpr} since the previous work~\cite{wei2019mmgcn} has validated the superior performance of MMGCN over these multi-modal recommender models. 

\vspace{5pt}
\noindent\textbf{Parameter Settings.}
We strictly follow the original implementation of MMGCN~\cite{wei2019mmgcn}, including code, parameter initialization, and hyperparameter tuning. The additional weight $\alpha$ in the multi-task loss function is tuned in $\{0, 0.25, 0.5, 0.75, 1, 2, 3, 4, 5\}$. The effect of $\alpha$ on the performance is visualized in Figure \ref{fig:alpha} where the model obtains the best performance when $\alpha$ is 1 or 2, showing the effectiveness of our proposed multi-task training. As shown in Table \ref{tab:effect_T}, we estimate the NDE of $E=e$ on $Y$ under situation $T=t^*$ due to its rationality and better performance. 
Moreover, early stopping is performed for the model selection, \ie stop training if recall@10 on the validation set does not increase for 10 successive epochs. We train all the models multiple times and report the average performance. 
More details can be found in the code\footnote{\url{https://github.com/WenjieWWJ/Clickbait/.}}.

\begin{table}[t]
\vspace{-3pt}
\setlength{\abovecaptionskip}{0cm}
\setlength{\belowcaptionskip}{0cm}
\caption{Results of estimating NDE \wrt different situations.}
\small
\label{tab:effect_T}
\begin{tabular}{l|l|l|l|l}
\hline
\textbf{} & \multicolumn{2}{c|}{\textbf{Tiktok}} & \multicolumn{2}{c}{\textbf{Adressa}} \\ \hline
\textbf{Method} & \textbf{R@20} & \textbf{N@20} & \textbf{R@20} & \textbf{N@20} \\ \hline
\textbf{NT} & 0.0635 & 0.0430 & 0.1612 & 0.1059 \\ \hline
\textbf{CR ($T=t$)} & 0.0671 & 0.0465 & 0.1667 & 0.1093 \\ \hline
\textbf{CR ($T=t^*$)} & \textbf{0.0683} & \textbf{0.0476} & \textbf{0.1712} & \textbf{0.1133} \\ \hline
\end{tabular}
\vspace{-0cm}
\end{table}
\begin{figure}[tb]
\centering
\vspace{-0.3cm}
\setlength{\abovecaptionskip}{0cm}
\setlength{\belowcaptionskip}{-0.4cm}
\includegraphics[scale=0.2]{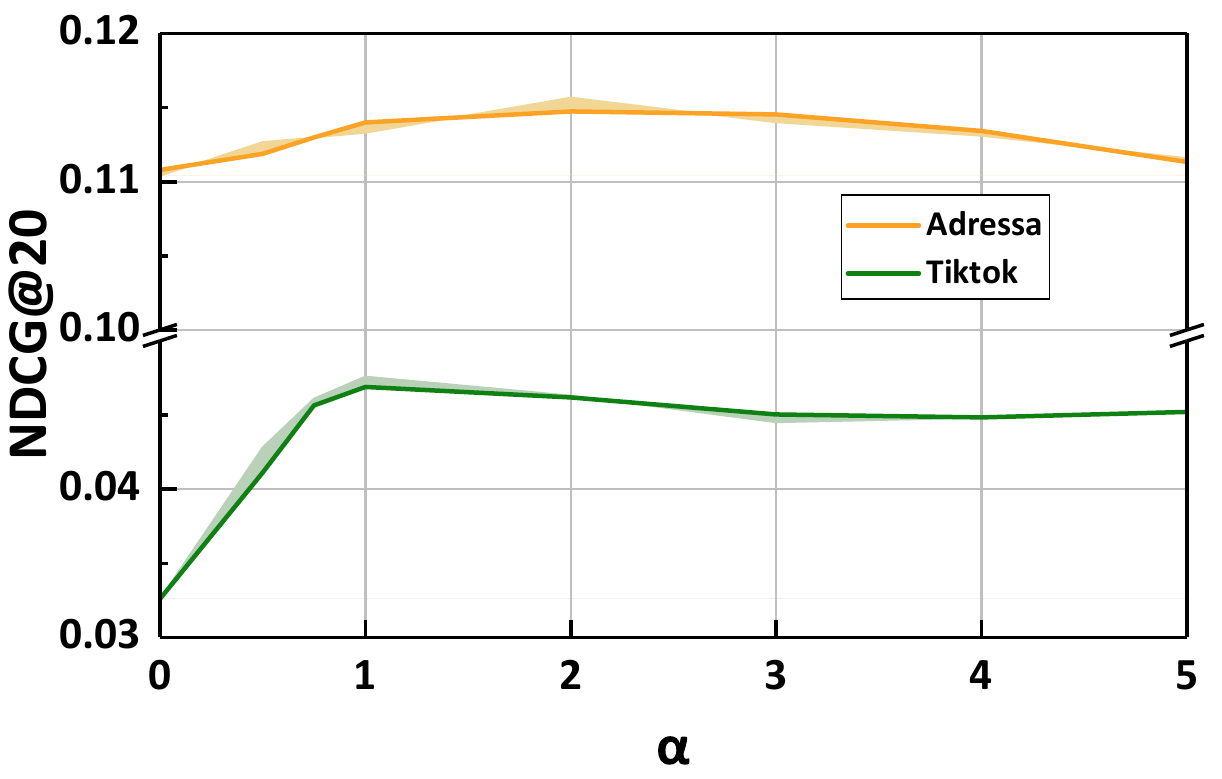}
\caption{Effect of $\alpha$ in the multi-task loss.}
\label{fig:alpha}
\end{figure}

\subsection{Performance Comparison}\label{ssec:perf_comp}
The overall performance comparison is summarized in Table \ref{tab:Results}. From the table, we have the following observations:


\vspace{5pt}
\noindent$\bullet$ \textbf{Debiasing Training}. 
In most cases, CFT performs worse than NT, which is attributed to discarding exposure features. The result overrules the option of simply discarding exposure features to mitigate the clickbait issue, which is indispensable for user preference prediction. Moreover, the performance of IPW is inferior on Tiktok and Adressa, showing that the clickbait issue may not be resolved by simply discouraging the recommendation of items with more clicks. In addition, the result indicates the importance of accurate propensity estimation to mitigate a bias, which is the crucial barrier of the usage of IPW for handling the bias caused by features with complex and changeable patterns. 

\vspace{5pt}
\noindent$\bullet$ \textbf{Post-click Feedback}. 
RR outperforms NT, which re-ranks the recommendations of NT according to the like/click ratio. It validates the effectiveness of leveraging post-click feedback to mitigate the clickbait issue and satisfy user requirements. However, CT and NR, which incorporate post-click feedback into the model training, perform worse than NT on Tiktok, \eg the NDCG@10 of CT decreases by 11.71\% on Tiktok. 
We ascribe the inferior performance to the sparsity of post-click feedback, which hurts the model generalization when the model is trained on a small number of interactions. 
It makes sense since the clicks that end with likes in Tiktok account for only $39.44\%$, which is much lower than that in Adressa $(69.47\%)$.
Moreover, we postulate the reason to be the inaccurate causal graph (Figure~\ref{fig:causal_our}(a)) that lacks the direct edge from exposure features to prediction, which is further detailed in Table~\ref{tab:TEforInference}.


\vspace{5pt}
\noindent$\bullet$ \textbf{CR Inference}. 
In all cases, CR achieves significant performance gains over all baselines. In particular, CR outperforms NT \wrt N$@10$ by 11.11\% and 7.47\% on Tiktok and Adressa, respectively. The result validates the effectiveness of the proposed CR, which is attributed to the new causal graph and counterfactual inference. In particular, CR also outperforms RR which additionally considers the post-click feedback. This further signifies the rationality of CR in eliminating the direct effect of exposure features on the prediction to mitigate the clickbait issue. As such, CR significantly helps to recommend more satisfying items, which can improve the user engagement and produce greater economic benefits.


%
%
%
%
%


\begin{table}[t]
\setlength{\abovecaptionskip}{0cm}
\setlength{\belowcaptionskip}{0cm}
\caption{Performance comparison between CR inference and the inference via TE.}\label{tab:TEforInference}
\begin{center}
\resizebox{0.45\textwidth}{!}{
\begin{tabular}{l|ccc|ccc}
\hline
\textbf{Dataset} & \multicolumn{3}{c|}{\textbf{Tiktok}} & \multicolumn{3}{c}{\textbf{Adressa}} \\
\textbf{Metric} & \textbf{P@20} & \textbf{R@20} & \textbf{N@20} & \textbf{P@20} & \textbf{R@20} & \textbf{N@20} \\ \hline \hline 
\textbf{NT}  & {0.0231} & {{0.0635}}  & 0.0430 & 0.0415 & {{0.1612}} & {{0.1059}}\\ \hline
\textbf{CR-TE} & {0.0235} & 0.0665 & 0.0461 & {0.0436} & 0.1698 & 0.1122 \\
\textbf{CR inference} & \textbf{0.0242} & \textbf{0.0683} & \textbf{0.0476} & \textbf{0.0439} & \textbf{0.1712} & \textbf{0.1133} \\\hline
\end{tabular}
}
\end{center}
\vspace{-0.4cm}
\end{table}

\begin{figure}[tb]
\centering
\setlength{\abovecaptionskip}{0.1cm}
\setlength{\belowcaptionskip}{-0.5cm}
\includegraphics[scale=0.2]{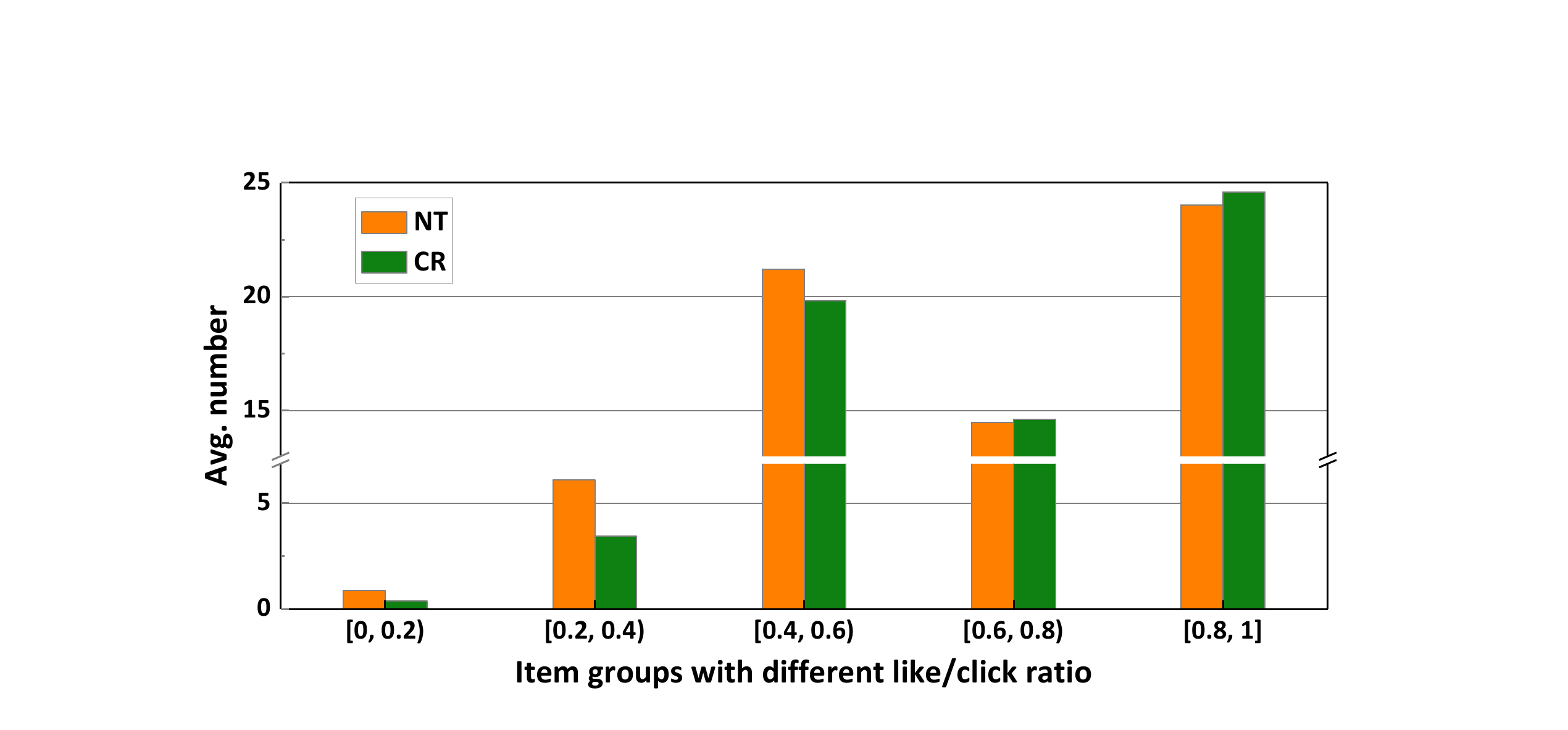}
\caption{Visualization of the averaged recommendation frequencies of items. Note that items with low like/click ratios shouldn't be recommended.}
\label{fig:groupLikeClick}
\end{figure}

\subsubsection{\textbf{Effect of the Proposed Causal Graph.}}
To shed light on the performance gain, we further study one variant, \ie CR-TE, which performs inference via the TE of $E=e$ and $T=t$, \ie its difference from NT is training over the proposed causal graph. Table~\ref{tab:TEforInference} shows their performance with $K=20$. From the table, we observe that CR-TE outperforms NT, which justifies the rationality of incorporating the direct edge from exposure features to the prediction score. It validates the existence of the shortcut where exposure features can directly lead to clicks.
Moreover, CR inference further outperforms CR-TE, showing that reducing the direct effect of exposure features indeed mitigates the clickbait issue and leads to better recommendation with more satisfaction.


\subsection{In-depth Analysis}
We then take CR on Adressa as an example to further investigate the effectiveness of CR. 
\subsubsection{\textbf{Visualization of Recommendations \wrt Like/click Ratio.}}
Recall that recommender models with the clickbait issue tend to recommend items even though their like/click ratios are low. 
We thus compare the recommendations of CR and NT to explore whether CR can reduce recommending the items with high risk to hurt user experience. Specifically, we collect top-ranked
items recommended to each user and count the frequency of each item being recommended. Figure \ref{fig:groupLikeClick} outlines the recommendation frequencies of CR and NT where items are intuitively split into five groups according to their like/click ratio for better visualization.
From the figure, we can see that as compared to NT, 1) CR recommends fewer items with like/click ratios $\leq 0.6$; and 2) more items with high like/click ratios, especially in $[0.8, 1]$. 
The result indicates the higher potential of CR to satisfy users, which is attributed to the proper modeling of the effect of exposure features.

\subsubsection{\textbf{Effect of Dataset Cleanness.}}
\begin{figure}[tb]
\centering
\setlength{\abovecaptionskip}{0cm}
\setlength{\belowcaptionskip}{0cm}
\includegraphics[scale=0.42]{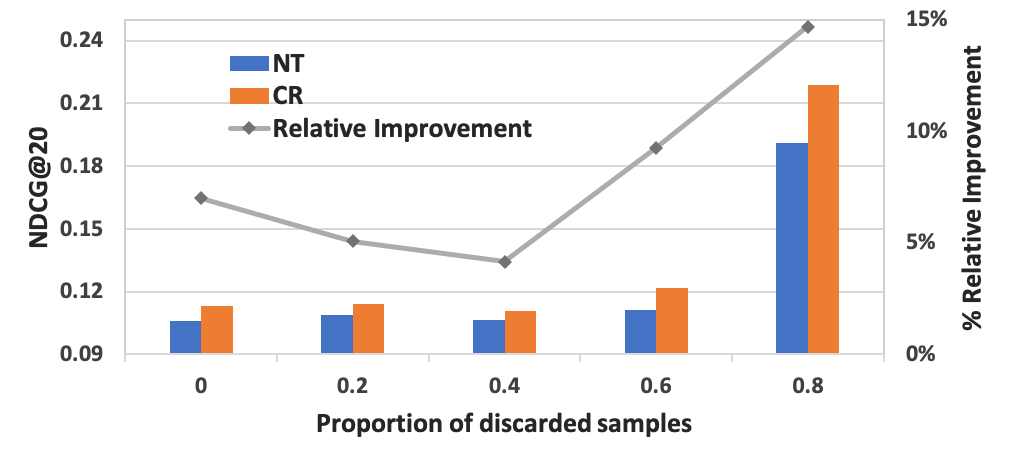}
\caption{Performance comparison across the subsets of Adressa with different discarding proportions. A larger proportion indicates a higher percentage of the clicks that end with dislikes in the dataset.}
\label{fig:splitTest}
\vspace{-0.3cm}
\end{figure}

We then study how the effectiveness of CR is influenced by the ``cleanness'' of the click data. Specifically, we compare CR and NT over filtered datasets with different percentages of clicks that end with dislikes. We rank the items in descending order by the like/click ratio, and discard the top-ranked items at a certain proportion where a larger discarding proportion leads to a dataset with a higher percentage of clicks that end with dislikes. Figure \ref{fig:splitTest} shows the performance with discarding proportion changing from 0 (the original dataset) to 0.8.
%
From Figure \ref{fig:splitTest}, we have the following findings: 
1) CR outperforms NT in all cases, which further validates the effectiveness of CR.
2) The performance gains are close when the discarding proportion is smaller than $0.4$, and increase dramatically under larger proportions. The result indicates that mitigating the clickbait issue is more important for the recommendation scenarios with more clicks that end with dislikes.


\subsubsection{\textbf{Effect of Fusion Strategy.}}
\label{sec:fusion_strategy}
Recall that any differentiable arithmetic binary operations can be equipped as the fusion strategy in CR~\cite{niu2020counterfactual}. To shed light on the development of proper fusion strategies, we investigate its essential properties, such as linearity and boundary.
As such, in addition to the MUL strategy, we further evaluate a vanilla SUM strategy with linear fusion, SUM with sigmoid function, and SUM/MUL with $tanh(\cdot)$ as the activation function. Formally, 
\begin{equation}\small
\left\{
\begin{aligned}
&\text{SUM-linear: } Y_{u, i, e} = f(Y_{u, i}, Y_{u, e}) = Y_{u, i} + Y_{u, e},\\
&\text{SUM-sigmoid: } Y_{u, i, e} = f(Y_{u, i}, Y_{u, e}) = Y_{u, i} + \sigma(Y_{u, e}),\\
&\text{SUM-tanh: }Y_{u, i, e} = f(Y_{u, i}, Y_{u, e}) = Y_{u, i} + \tanh(Y_{u, e}),\\
&\text{MUL-tanh: }Y_{u, i, e} = f(Y_{u, i}, Y_{u, e}) = Y_{u, i} * \tanh(Y_{u, e}).
\end{aligned}
\right .
\end{equation}
Similar to the MUL fusion strategy, we also estimate CR inference for SUM-linear, SUM-sigmoid, SUM-tanh, and MUL-tanh, respectively. The results are as follows:
\begin{equation}\small
\notag
\left \{
\begin{aligned}
&\text{SUM-linear: } Y_{CR} = Y_{u, i} - c_{u, i} \propto Y_{u, i},\\ 
&\text{SUM-sigmoid: } Y_{CR} = Y_{u, i} - c_{u, i} \propto Y_{u, i},\\ 
&\text{SUM-tanh: } Y_{CR} = Y_{u, i} - c_{u, i} \propto Y_{u, i},\\ 
&\text{MUL-tanh: } Y_{CR} = (Y_{u, i} - c_{u, i}) * \tanh(Y_{u, e}).\\ 
\end{aligned}
\right .
\end{equation}
During CR inference, the SUM strategies with different activation functions are equivalent. However, they capture the direct effect of exposure features differently in the training process. Therefore, the recommendation results are theoretically different. 

The performance of different fusion strategies is reported in Table \ref{tab:FusionStrategy}. From that, we can find that: 1) non-linear fusion strategies are significantly better than linear ones due to the better representation capacity; and 2) SUM-tanh achieves the best performance over the other fusion strategies, including the proposed MUL-sigmoid strategy. This shows that a fusion function with the proper boundary can further improve the performance of CR and multiple fusion strategies are worth studying when CR inference is applied to other datasets in future.

\begin{table}[tbp]
\setlength{\abovecaptionskip}{0cm}
\setlength{\belowcaptionskip}{0cm}
\caption{Performance of CR with different fusion strategies.}\label{tab:FusionStrategy}
\begin{center}
\resizebox{0.47\textwidth}{!}{
\begin{tabular}{l|ccc|ccc}
\hline
\textbf{Metric} & \textbf{P@10} & \textbf{R@10} & \textbf{N@10} & \textbf{P@20}& \textbf{R@20} & \textbf{N@20} \\ \hline  
\textbf{SUM-Linear} & 0.0380 & 0.0718 & 0.0598 & 0.0317 & 0.1196 & 0.0780 \\\hline
\textbf{SUM-tanh} & \textbf{0.0537} & \textbf{0.1060} & \textbf{0.0889} & \textbf{0.0447} & \textbf{0.1744} & \textbf{0.1150} \\
\textbf{MUL-tanh} & 0.0520 & 0.1027 & 0.0861 & 0.0435 & 0.1698 & 0.1118 \\\hline
\textbf{SUM-sigmoid} & 0.0533 & 0.1044 & 0.0877 & 0.0439 & 0.1714 & 0.1132 \\
\textbf{MUL-sigmoid} & 0.0532 & 0.1045 & 0.0878 & 0.0439 & 0.1711 & 0.1132\\\hline
\end{tabular}
}
\end{center}
\vspace{-0.3cm}
\end{table}




\subsubsection{\textbf{CR Evaluation on Synthetic Data}}
To further evaluate the effectiveness of CR on mitigating the direct effect of exposure features, we conduct experiments on synthetic data. Specifically, during inference, we construct a fake item for each positive user-item pair in the testing data by ``poisoning'' the exposure feature of the item. The content features of the fake item are the same as the real item while its exposure features are randomly selected from the items with the like/click ratio < 0.5. 
Such items with low like/click ratios are more likely to be the ones with the clickbait issue. Their exposure features are easy to be attractive but deceptive, for example, ``Find UFO!''. 
Besides, there is a large discrepancy between the exposure and content features of the fake items, which simulates the items with the clickbait issue where content features do not align with exposure features.
Therefore, the fake item should have a lower rank than the paired real item if the recommender model can mitigate the clickbait issue well. 


A lower rank of the fake item indicates a better elimination of the direct effect from the exposure features.
Accordingly, we rank all testing real items and the fakes ones for each user, and we define $\textit{rank\_gap} = rank_{fake} - rank_{real}$ to measure the performance of recommender models, where $rank_{fake}$ and $rank_{real}$ are the ranks of the paired fake and real items, respectively. 
A larger $\textit{rank\_gap}$ value indicates a bigger gap and thus better performance. Lastly, we calculate the $\textit{rank\_gap}$ of each triplet <user, real item, fake item> in the testing data. 

\begin{figure}[t]
\setlength{\abovecaptionskip}{0.1cm}
\setlength{\belowcaptionskip}{-0.4cm}
  \centering 
  \hspace{-0.2in}
    \subfigure[Distribution \wrt $\textit{rank\_gap}$ group.]{
    \label{fig:synthetic:a}
    \includegraphics[width=2in]{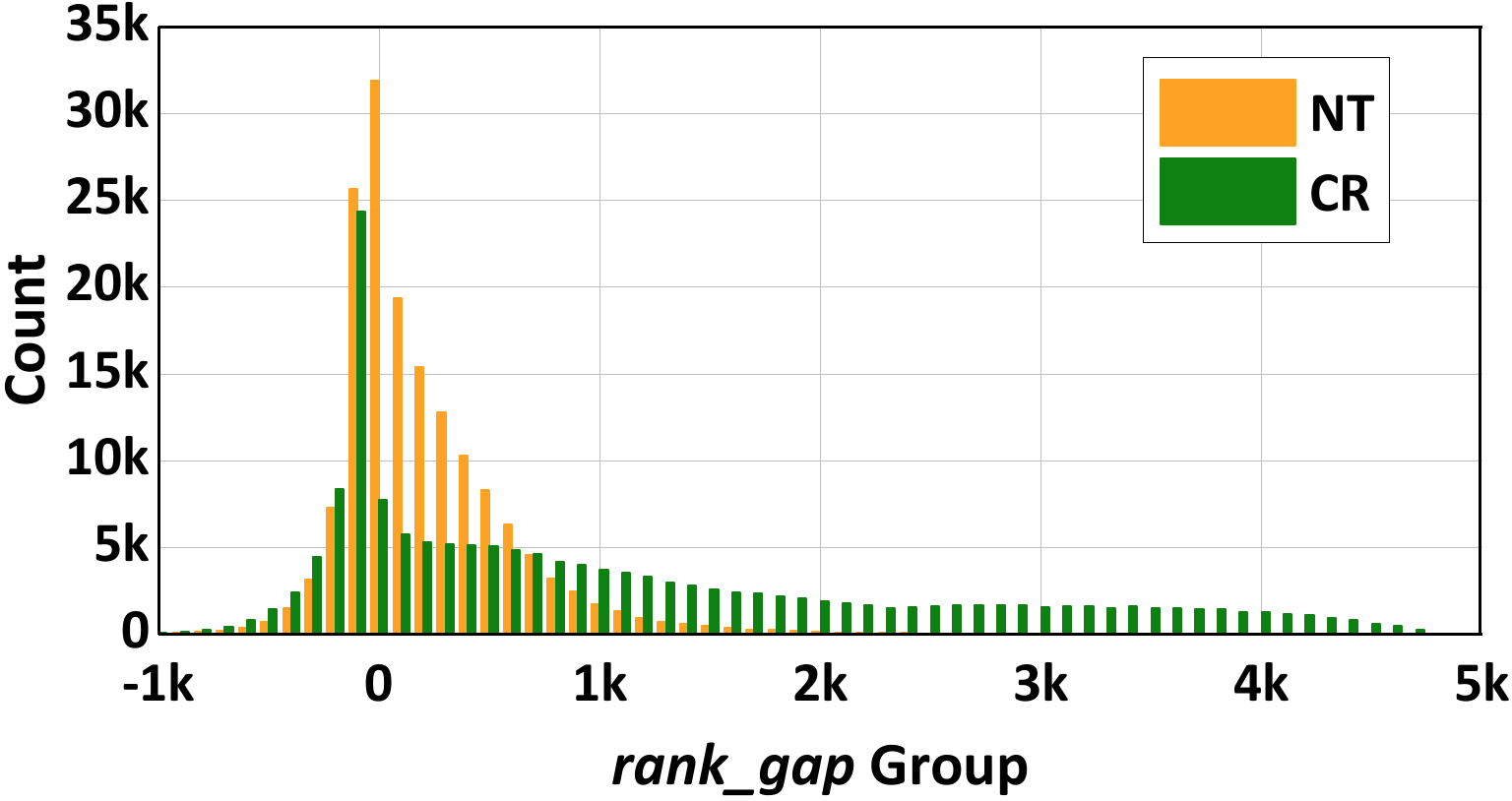}}
    \subfigure[$\textit{rank\_gap}$ of NT and CR.]{
    \includegraphics[width=1.3in]{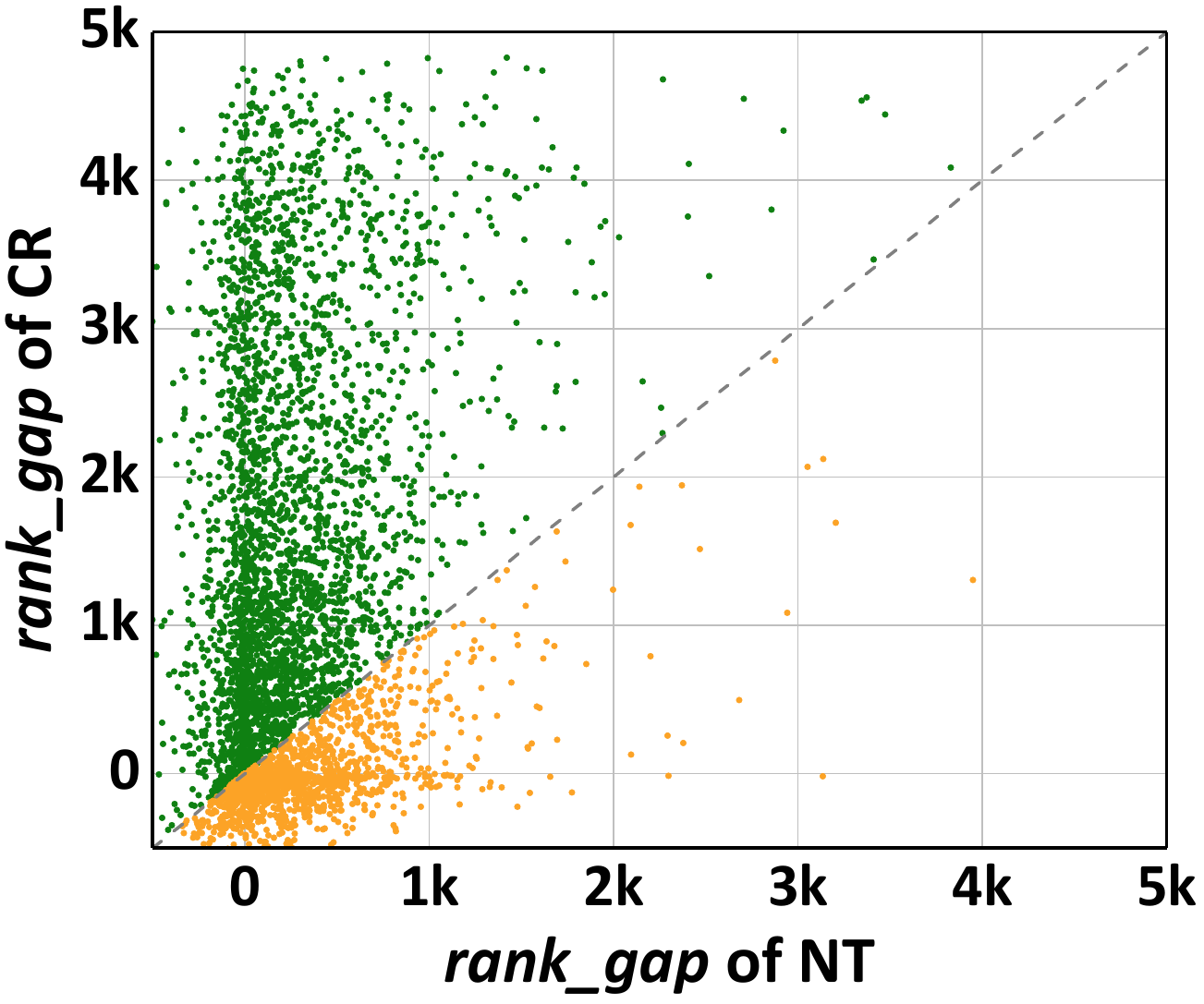}
    \label{fig:synthetic:b}
    }
    \caption{Results of CR evaluation on synthetic data.}
      \label{fig:synthetic}
\end{figure}

As shown in Figure \ref{fig:synthetic:a}, the $\textit{rank\_gap}$ values are first grouped, and then counted by group. 
From this figure, we can observe that the $\textit{rank\_gap}$ values generated by CR are larger than those of NT, and the distribution of CR is flatter than that of NT, indicating that CR produces lower ranking scores for the fake items. This is because CR effectively reduces the direct effect of deceptive exposure features. Besides, we randomly sample 5k samples of triplets from the testing data and individually compare the $\textit{rank\_gap}$ values generated by CR and NT in Figure \ref{fig:synthetic:b}. From the figure, we can find that 1) most points are above the diagonal, showing the $\textit{rank\_gap}$ of CR is usually larger than that of NT; and 2) the $\textit{rank\_gap}$ values generated by CR cover a wider range, varying from 0 to 5k. The findings imply that CR can distinguish the real and fake items well, which further proves the effectiveness of CR on mitigating the clickbait issue.

\section{Conclusion and Future Work}
\label{sec:discussion}

The clickbait issue widely exists in the industrial recommender systems. To eliminate its effect, we proposed a new recommendation framework CR that accounts for the causal relations among the exposure features, content features, and predictions. Through performing counterfactual inference, we estimated the direct effect of exposure features on the prediction and removed it from recommendation scoring. While we instantiated CR on a specific recommender model MMGCN, it is model-agnostic and only requires minor adjustments (several lines of codes) to be adopted to other models, enabling the wide usage of CR across different recommendation scenarios and models. By mitigating the clickbait issue, they can improve the user satisfaction and engagement.

This work opens up a new research direction---incorporating counterfactual inference into recommender systems.
Following this direction, there are many interesting ideas that deserve our exploration. 
1) Considering the huge benefit of reasoning over causal graph, it is essential to construct a more comprehensive causal graph for recommendation with more fine-grained causal relations in future.
2) This work justifies the effectiveness of counterfactual inference on mitigating the clickbait issue, and motivates further exploration on other intrinsic biases and issues in the click data, such as selection bias \cite{Ovaisi2020Correcting} and position bias~\cite{Thorsten2017unbiased}.
3) More broadly, this work signifies the importance of causal inference on recommendation. It opens the door of empowering recommender systems with more causal inference techniques, such as intervention and counterfactual inference. 

\begin{acks}
This research/project is supported by the Sea-NExT Joint Lab, the National Natural Science Foundation of China (U19A2079) and National Key Research and Development Program of China (2020AAA0106000).
\end{acks}

{
\tiny
\bibliographystyle{ACM-Reference-Format}
\balance
\bibliography{bibfile}


\begin{thebibliography}{60}


\ifx \showCODEN    \undefined \def \showCODEN     #1{\unskip}     \fi
\ifx \showDOI      \undefined \def \showDOI       #1{#1}\fi
\ifx \showISBNx    \undefined \def \showISBNx     #1{\unskip}     \fi
\ifx \showISBNxiii \undefined \def \showISBNxiii  #1{\unskip}     \fi
\ifx \showISSN     \undefined \def \showISSN      #1{\unskip}     \fi
\ifx \showLCCN     \undefined \def \showLCCN      #1{\unskip}     \fi
\ifx \shownote     \undefined \def \shownote      #1{#1}          \fi
\ifx \showarticletitle \undefined \def \showarticletitle #1{#1}   \fi
\ifx \showURL      \undefined \def \showURL       {\relax}        \fi
\providecommand\bibfield[2]{#2}
\providecommand\bibinfo[2]{#2}
\providecommand\natexlab[1]{#1}
\providecommand\showeprint[2][]{arXiv:#2}

\bibitem[\protect\citeauthoryear{Abdollahpouri, Burke, and
  Mobasher}{Abdollahpouri et~al\mbox{.}}{2019}]%
        {abdollahpouri2019managing}
\bibfield{author}{\bibinfo{person}{Himan Abdollahpouri}, \bibinfo{person}{Robin
  Burke}, {and} \bibinfo{person}{Bamshad Mobasher}.}
  \bibinfo{year}{2019}\natexlab{}.
\newblock \showarticletitle{Managing popularity bias in recommender systems
  with personalized re-ranking}. In \bibinfo{booktitle}{{\em Proceedings of the
  International Flairs Conference}}. \bibinfo{publisher}{AAAI Press}.
\newblock


\bibitem[\protect\citeauthoryear{Ai, Bi, Luo, Guo, and Croft}{Ai
  et~al\mbox{.}}{2018}]%
        {ai2018unbiased}
\bibfield{author}{\bibinfo{person}{Qingyao Ai}, \bibinfo{person}{Keping Bi},
  \bibinfo{person}{Cheng Luo}, \bibinfo{person}{Jiafeng Guo}, {and}
  \bibinfo{person}{W.~Bruce Croft}.} \bibinfo{year}{2018}\natexlab{}.
\newblock \showarticletitle{Unbiased Learning to Rank with Unbiased Propensity
  Estimation}. In \bibinfo{booktitle}{{\em SIGIR}}. \bibinfo{publisher}{ACM},
  \bibinfo{pages}{385--394}.
\newblock


\bibitem[\protect\citeauthoryear{Bonner and Vasile}{Bonner and Vasile}{2018}]%
        {bonner2018causal}
\bibfield{author}{\bibinfo{person}{Stephen Bonner} {and}
  \bibinfo{person}{Flavian Vasile}.} \bibinfo{year}{2018}\natexlab{}.
\newblock \showarticletitle{Causal embeddings for recommendation}. In
  \bibinfo{booktitle}{{\em RecSys}}. \bibinfo{publisher}{ACM},
  \bibinfo{pages}{104--112}.
\newblock


\bibitem[\protect\citeauthoryear{Cadene, Dancette, Ben-younes, Cord, Parikh,
  et~al\mbox{.}}{Cadene et~al\mbox{.}}{2019}]%
        {cadene2019rubi}
\bibfield{author}{\bibinfo{person}{Remi Cadene}, \bibinfo{person}{Corentin
  Dancette}, \bibinfo{person}{Hedi Ben-younes}, \bibinfo{person}{Matthieu
  Cord}, \bibinfo{person}{Devi Parikh}, {et~al\mbox{.}}}
  \bibinfo{year}{2019}\natexlab{}.
\newblock \showarticletitle{Rubi: Reducing unimodal biases for visual question
  answering}. In \bibinfo{booktitle}{{\em NeuIPS}}. \bibinfo{pages}{841--852}.
\newblock


\bibitem[\protect\citeauthoryear{Chen, Dong, Wang, Feng, Wang, and He}{Chen
  et~al\mbox{.}}{2020}]%
        {chen2020bias}
\bibfield{author}{\bibinfo{person}{Jiawei Chen}, \bibinfo{person}{Hande Dong},
  \bibinfo{person}{Xiang Wang}, \bibinfo{person}{Fuli Feng},
  \bibinfo{person}{Meng Wang}, {and} \bibinfo{person}{Xiangnan He}.}
  \bibinfo{year}{2020}\natexlab{}.
\newblock \showarticletitle{Bias and Debias in Recommender System: A Survey and
  Future Directions}.
\newblock \bibinfo{journal}{{\em arXiv preprint arXiv:2010.03240\/}}
  (\bibinfo{year}{2020}).
\newblock


\bibitem[\protect\citeauthoryear{Chen, Zhang, He, Nie, Liu, and Chua}{Chen
  et~al\mbox{.}}{2017}]%
        {chen2017attentive}
\bibfield{author}{\bibinfo{person}{Jingyuan Chen}, \bibinfo{person}{Hanwang
  Zhang}, \bibinfo{person}{Xiangnan He}, \bibinfo{person}{Liqiang Nie},
  \bibinfo{person}{Wei Liu}, {and} \bibinfo{person}{Tat-Seng Chua}.}
  \bibinfo{year}{2017}\natexlab{}.
\newblock \showarticletitle{Attentive collaborative filtering: Multimedia
  recommendation with item-and component-level attention}. In
  \bibinfo{booktitle}{{\em SIGIR}}. \bibinfo{publisher}{ACM},
  \bibinfo{pages}{335--344}.
\newblock


\bibitem[\protect\citeauthoryear{Chen, Liu, Zha, Zhou, Xiong, and Li}{Chen
  et~al\mbox{.}}{2018}]%
        {Chen2018temporal}
\bibfield{author}{\bibinfo{person}{Xusong Chen}, \bibinfo{person}{Dong Liu},
  \bibinfo{person}{Zheng-Jun Zha}, \bibinfo{person}{Wengang Zhou},
  \bibinfo{person}{Zhiwei Xiong}, {and} \bibinfo{person}{Yan Li}.}
  \bibinfo{year}{2018}\natexlab{}.
\newblock \showarticletitle{Temporal Hierarchical Attention at Category- and
  Item-Level for Micro-Video Click-Through Prediction}. In
  \bibinfo{booktitle}{{\em MM}}. \bibinfo{publisher}{ACM},
  \bibinfo{pages}{1146--1153}.
\newblock


\bibitem[\protect\citeauthoryear{Christakopoulou, Traverse, Potter, Marriott,
  Li, Haulk, Chi, and Chen}{Christakopoulou et~al\mbox{.}}{2020}]%
        {Konstantina2020Deconfounding}
\bibfield{author}{\bibinfo{person}{Konstantina Christakopoulou},
  \bibinfo{person}{Madeleine Traverse}, \bibinfo{person}{Trevor Potter},
  \bibinfo{person}{Emma Marriott}, \bibinfo{person}{Daniel Li},
  \bibinfo{person}{Chris Haulk}, \bibinfo{person}{Ed~H. Chi}, {and}
  \bibinfo{person}{Minmin Chen}.} \bibinfo{year}{2020}\natexlab{}.
\newblock \showarticletitle{Deconfounding User Satisfaction Estimation from
  Response Rate Bias}. In \bibinfo{booktitle}{{\em RecSys}}.
  \bibinfo{publisher}{ACM}, \bibinfo{pages}{450--455}.
\newblock


\bibitem[\protect\citeauthoryear{Craswell, Zoeter, Taylor, and Ramsey}{Craswell
  et~al\mbox{.}}{2008}]%
        {Nick2008an}
\bibfield{author}{\bibinfo{person}{Nick Craswell}, \bibinfo{person}{Onno
  Zoeter}, \bibinfo{person}{Michael Taylor}, {and} \bibinfo{person}{Bill
  Ramsey}.} \bibinfo{year}{2008}\natexlab{}.
\newblock \showarticletitle{An Experimental Comparison of Click Position-Bias
  Models}. In \bibinfo{booktitle}{{\em WSDM}}. \bibinfo{publisher}{ACM},
  \bibinfo{pages}{87--94}.
\newblock


\bibitem[\protect\citeauthoryear{Cs{\'a}ji}{Cs{\'a}ji}{2001}]%
        {csaji2001approximation}
\bibfield{author}{\bibinfo{person}{Bal{\'a}zs~Csan{\'a}d Cs{\'a}ji}.}
  \bibinfo{year}{2001}\natexlab{}.
\newblock \showarticletitle{Approximation with artificial neural networks}.
\newblock \bibinfo{journal}{{\em Faculty of Sciences, Etvs Lornd University,
  Hungary\/}} \bibinfo{volume}{24}, \bibinfo{number}{48}
  (\bibinfo{year}{2001}), \bibinfo{pages}{7}.
\newblock


\bibitem[\protect\citeauthoryear{Devlin, Chang, Lee, and Toutanova}{Devlin
  et~al\mbox{.}}{2018}]%
        {devlin2018bert}
\bibfield{author}{\bibinfo{person}{Jacob Devlin}, \bibinfo{person}{Ming-Wei
  Chang}, \bibinfo{person}{Kenton Lee}, {and} \bibinfo{person}{Kristina
  Toutanova}.} \bibinfo{year}{2018}\natexlab{}.
\newblock \showarticletitle{\text{BERT}: Pre-training of Deep Bidirectional
  Transformers for Language Understanding}. In \bibinfo{booktitle}{{\em
  arXiv:1810.04805}}.
\newblock


\bibitem[\protect\citeauthoryear{Fan, Ma, Li, He, Zhao, Tang, and Yin}{Fan
  et~al\mbox{.}}{2019}]%
        {fan2019graph}
\bibfield{author}{\bibinfo{person}{Wenqi Fan}, \bibinfo{person}{Yao Ma},
  \bibinfo{person}{Qing Li}, \bibinfo{person}{Yuan He}, \bibinfo{person}{Eric
  Zhao}, \bibinfo{person}{Jiliang Tang}, {and} \bibinfo{person}{Dawei Yin}.}
  \bibinfo{year}{2019}\natexlab{}.
\newblock \showarticletitle{Graph Neural Networks for Social Recommendation}.
  In \bibinfo{booktitle}{{\em WWW}}. \bibinfo{publisher}{ACM},
  \bibinfo{pages}{417--426}.
\newblock


\bibitem[\protect\citeauthoryear{Feng, He, Liu, Nie, and Chua}{Feng
  et~al\mbox{.}}{2018}]%
        {feng2018learning}
\bibfield{author}{\bibinfo{person}{Fuli Feng}, \bibinfo{person}{Xiangnan He},
  \bibinfo{person}{Yiqun Liu}, \bibinfo{person}{Liqiang Nie}, {and}
  \bibinfo{person}{Tat-Seng Chua}.} \bibinfo{year}{2018}\natexlab{}.
\newblock \showarticletitle{Learning on partial-order hypergraphs}. In
  \bibinfo{booktitle}{{\em WWW}}. \bibinfo{publisher}{ACM},
  \bibinfo{pages}{1523--1532}.
\newblock


\bibitem[\protect\citeauthoryear{Feng, He, Tang, and Chua}{Feng
  et~al\mbox{.}}{2019}]%
        {feng2019graph}
\bibfield{author}{\bibinfo{person}{Fuli Feng}, \bibinfo{person}{Xiangnan He},
  \bibinfo{person}{Jie Tang}, {and} \bibinfo{person}{Tat-Seng Chua}.}
  \bibinfo{year}{2019}\natexlab{}.
\newblock \showarticletitle{Graph adversarial training: Dynamically
  regularizing based on graph structure}.
\newblock \bibinfo{journal}{{\em TKDE\/}} (\bibinfo{year}{2019}).
\newblock


\bibitem[\protect\citeauthoryear{Goodfellow, Bengio, and Courville}{Goodfellow
  et~al\mbox{.}}{2016}]%
        {Goodfellow2016deep}
\bibfield{author}{\bibinfo{person}{Ian Goodfellow}, \bibinfo{person}{Yoshua
  Bengio}, {and} \bibinfo{person}{Aaron Courville}.}
  \bibinfo{year}{2016}\natexlab{}.
\newblock \bibinfo{booktitle}{{\em Deep Learning}}.
\newblock \bibinfo{publisher}{MIT Press}.
\newblock


\bibitem[\protect\citeauthoryear{Gulla, Zhang, Liu, \"{O}zg\"{o}bek, and
  Su}{Gulla et~al\mbox{.}}{2017}]%
        {Gulla2017the}
\bibfield{author}{\bibinfo{person}{Jon~Atle Gulla}, \bibinfo{person}{Lemei
  Zhang}, \bibinfo{person}{Peng Liu}, \bibinfo{person}{\"{O}zlem
  \"{O}zg\"{o}bek}, {and} \bibinfo{person}{Xiaomeng Su}.}
  \bibinfo{year}{2017}\natexlab{}.
\newblock \showarticletitle{The Adressa Dataset for News Recommendation}. In
  \bibinfo{booktitle}{{\em WI}}. \bibinfo{publisher}{ACM},
  \bibinfo{pages}{1042--1048}.
\newblock


\bibitem[\protect\citeauthoryear{He and McAuley}{He and McAuley}{2016}]%
        {he2016vbpr}
\bibfield{author}{\bibinfo{person}{Ruining He} {and} \bibinfo{person}{Julian
  McAuley}.} \bibinfo{year}{2016}\natexlab{}.
\newblock \showarticletitle{\text{VBPR}: visual bayesian personalized ranking
  from implicit feedback}. In \bibinfo{booktitle}{{\em AAAI}}.
  \bibinfo{publisher}{AAAI press}.
\newblock


\bibitem[\protect\citeauthoryear{He, Deng, Wang, Li, Zhang, and Wang}{He
  et~al\mbox{.}}{2020}]%
        {LightGCN2020he}
\bibfield{author}{\bibinfo{person}{Xiangnan He}, \bibinfo{person}{Kuan Deng},
  \bibinfo{person}{Xiang Wang}, \bibinfo{person}{Yan Li},
  \bibinfo{person}{YongDong Zhang}, {and} \bibinfo{person}{Meng Wang}.}
  \bibinfo{year}{2020}\natexlab{}.
\newblock \showarticletitle{LightGCN: Simplifying and Powering Graph
  Convolution Network for Recommendation}. In \bibinfo{booktitle}{{\em SIGIR}}.
  \bibinfo{publisher}{ACM}, \bibinfo{pages}{639–648}.
\newblock


\bibitem[\protect\citeauthoryear{He, Liao, Zhang, Nie, Hu, and Chua}{He
  et~al\mbox{.}}{2017}]%
        {He2017Neural}
\bibfield{author}{\bibinfo{person}{Xiangnan He}, \bibinfo{person}{Lizi Liao},
  \bibinfo{person}{Hanwang Zhang}, \bibinfo{person}{Liqiang Nie},
  \bibinfo{person}{Xia Hu}, {and} \bibinfo{person}{Tat-Seng Chua}.}
  \bibinfo{year}{2017}\natexlab{}.
\newblock \showarticletitle{Neural Collaborative Filtering}. In
  \bibinfo{booktitle}{{\em WWW}}. \bibinfo{publisher}{ACM},
  \bibinfo{pages}{173--182}.
\newblock


\bibitem[\protect\citeauthoryear{Hofmann, Behr, and Radlinski}{Hofmann
  et~al\mbox{.}}{2012}]%
        {Hofmann2012on}
\bibfield{author}{\bibinfo{person}{Katja Hofmann}, \bibinfo{person}{Fritz
  Behr}, {and} \bibinfo{person}{Filip Radlinski}.}
  \bibinfo{year}{2012}\natexlab{}.
\newblock \showarticletitle{On Caption Bias in Interleaving Experiments}. In
  \bibinfo{booktitle}{{\em CIKM}}. \bibinfo{publisher}{ACM},
  \bibinfo{pages}{115--124}.
\newblock


\bibitem[\protect\citeauthoryear{Hong, Li, Cai, Tao, Wang, and Tian}{Hong
  et~al\mbox{.}}{2017}]%
        {hong2017coherent}
\bibfield{author}{\bibinfo{person}{Richang Hong}, \bibinfo{person}{Lei Li},
  \bibinfo{person}{Junjie Cai}, \bibinfo{person}{Dapeng Tao},
  \bibinfo{person}{Meng Wang}, {and} \bibinfo{person}{Qi Tian}.}
  \bibinfo{year}{2017}\natexlab{}.
\newblock \showarticletitle{Coherent semantic-visual indexing for large-scale
  image retrieval in the cloud}.
\newblock \bibinfo{journal}{{\em TIP\/}} \bibinfo{volume}{26},
  \bibinfo{number}{9} (\bibinfo{year}{2017}), \bibinfo{pages}{4128--4138}.
\newblock


\bibitem[\protect\citeauthoryear{Hong, Yang, Wang, and Hua}{Hong
  et~al\mbox{.}}{2015}]%
        {hong2015learning}
\bibfield{author}{\bibinfo{person}{Richang Hong}, \bibinfo{person}{Yang Yang},
  \bibinfo{person}{Meng Wang}, {and} \bibinfo{person}{Xian-Sheng Hua}.}
  \bibinfo{year}{2015}\natexlab{}.
\newblock \showarticletitle{Learning visual semantic relationships for
  efficient visual retrieval}.
\newblock \bibinfo{journal}{{\em Transactions on Big Data\/}}
  \bibinfo{volume}{1}, \bibinfo{number}{4} (\bibinfo{year}{2015}),
  \bibinfo{pages}{152--161}.
\newblock


\bibitem[\protect\citeauthoryear{Jiang, Wang, Wei, Gao, Wang, and Nie}{Jiang
  et~al\mbox{.}}{2020}]%
        {jiang2020aspect}
\bibfield{author}{\bibinfo{person}{Hao Jiang}, \bibinfo{person}{Wenjie Wang},
  \bibinfo{person}{Yinwei Wei}, \bibinfo{person}{Zan Gao},
  \bibinfo{person}{Yinglong Wang}, {and} \bibinfo{person}{Liqiang Nie}.}
  \bibinfo{year}{2020}\natexlab{}.
\newblock \showarticletitle{What Aspect Do You Like: Multi-scale Time-aware
  User Interest Modeling for Micro-video Recommendation}. In
  \bibinfo{booktitle}{{\em MM}}. \bibinfo{publisher}{ACM},
  \bibinfo{pages}{3487--3495}.
\newblock


\bibitem[\protect\citeauthoryear{Joachims, Swaminathan, and Schnabel}{Joachims
  et~al\mbox{.}}{2017}]%
        {Thorsten2017unbiased}
\bibfield{author}{\bibinfo{person}{Thorsten Joachims}, \bibinfo{person}{Adith
  Swaminathan}, {and} \bibinfo{person}{Tobias Schnabel}.}
  \bibinfo{year}{2017}\natexlab{}.
\newblock \showarticletitle{Unbiased Learning-to-Rank with Biased Feedback}. In
  \bibinfo{booktitle}{{\em WSDM}}. \bibinfo{publisher}{ACM},
  \bibinfo{pages}{781--789}.
\newblock


\bibitem[\protect\citeauthoryear{Kim, Hassan, White, and Zitouni}{Kim
  et~al\mbox{.}}{2014}]%
        {Kim2014Modeling}
\bibfield{author}{\bibinfo{person}{Youngho Kim}, \bibinfo{person}{Ahmed
  Hassan}, \bibinfo{person}{Ryen~W White}, {and} \bibinfo{person}{Imed
  Zitouni}.} \bibinfo{year}{2014}\natexlab{}.
\newblock \showarticletitle{Modeling dwell time to predict click-level
  satisfaction}. In \bibinfo{booktitle}{{\em WSDM}}. \bibinfo{publisher}{ACM},
  \bibinfo{pages}{193--202}.
\newblock


\bibitem[\protect\citeauthoryear{Li, Liu, Yin, Cui, Xu, and Nie}{Li
  et~al\mbox{.}}{2019}]%
        {li2019routing}
\bibfield{author}{\bibinfo{person}{Yongqi Li}, \bibinfo{person}{Meng Liu},
  \bibinfo{person}{Jianhua Yin}, \bibinfo{person}{Chaoran Cui},
  \bibinfo{person}{Xin-Shun Xu}, {and} \bibinfo{person}{Liqiang Nie}.}
  \bibinfo{year}{2019}\natexlab{}.
\newblock \showarticletitle{Routing micro-videos via a temporal graph-guided
  recommendation system}. In \bibinfo{booktitle}{{\em MM}}.
  \bibinfo{publisher}{ACM}, \bibinfo{pages}{1464--1472}.
\newblock


\bibitem[\protect\citeauthoryear{Liang, Charlin, and Blei}{Liang
  et~al\mbox{.}}{2016a}]%
        {liang2016causal}
\bibfield{author}{\bibinfo{person}{Dawen Liang}, \bibinfo{person}{Laurent
  Charlin}, {and} \bibinfo{person}{David~M Blei}.}
  \bibinfo{year}{2016}\natexlab{a}.
\newblock \showarticletitle{Causal inference for recommendation}. In
  \bibinfo{booktitle}{{\em UAI}}. AUAI.
\newblock


\bibitem[\protect\citeauthoryear{Liang, Charlin, McInerney, and Blei}{Liang
  et~al\mbox{.}}{2016b}]%
        {liang2016modeling}
\bibfield{author}{\bibinfo{person}{Dawen Liang}, \bibinfo{person}{Laurent
  Charlin}, \bibinfo{person}{James McInerney}, {and} \bibinfo{person}{David~M
  Blei}.} \bibinfo{year}{2016}\natexlab{b}.
\newblock \showarticletitle{Modeling user exposure in recommendation}. In
  \bibinfo{booktitle}{{\em WWW}}. \bibinfo{publisher}{ACM},
  \bibinfo{pages}{951--961}.
\newblock


\bibitem[\protect\citeauthoryear{Liu, White, and Dumais}{Liu
  et~al\mbox{.}}{2010}]%
        {liu2010understanding}
\bibfield{author}{\bibinfo{person}{Chao Liu}, \bibinfo{person}{Ryen~W White},
  {and} \bibinfo{person}{Susan Dumais}.} \bibinfo{year}{2010}\natexlab{}.
\newblock \showarticletitle{Understanding web browsing behaviors through
  Weibull analysis of dwell time}. In \bibinfo{booktitle}{{\em SIGIR}}. ACM,
  \bibinfo{pages}{379--386}.
\newblock


\bibitem[\protect\citeauthoryear{Liu, Cheng, Dong, He, Pan, and Ming}{Liu
  et~al\mbox{.}}{2020a}]%
        {Liu2020a}
\bibfield{author}{\bibinfo{person}{Dugang Liu}, \bibinfo{person}{Pengxiang
  Cheng}, \bibinfo{person}{Zhenhua Dong}, \bibinfo{person}{Xiuqiang He},
  \bibinfo{person}{Weike Pan}, {and} \bibinfo{person}{Zhong Ming}.}
  \bibinfo{year}{2020}\natexlab{a}.
\newblock \showarticletitle{A General Knowledge Distillation Framework for
  Counterfactual Recommendation via Uniform Data}. In \bibinfo{booktitle}{{\em
  SIGIR}}. \bibinfo{publisher}{ACM}, \bibinfo{pages}{831--840}.
\newblock


\bibitem[\protect\citeauthoryear{Liu, Qu, Nie, Liu, Duan, and Chen}{Liu
  et~al\mbox{.}}{2020b}]%
        {liu2020iterative}
\bibfield{author}{\bibinfo{person}{Meng Liu}, \bibinfo{person}{Leigang Qu},
  \bibinfo{person}{Liqiang Nie}, \bibinfo{person}{Maofu Liu},
  \bibinfo{person}{Lingyu Duan}, {and} \bibinfo{person}{Baoquan Chen}.}
  \bibinfo{year}{2020}\natexlab{b}.
\newblock \showarticletitle{Iterative Local-Global Collaboration Learning
  Towards One-Shot Video Person Re-Identification}.
\newblock \bibinfo{journal}{{\em TIP\/}}  \bibinfo{volume}{29}
  (\bibinfo{year}{2020}), \bibinfo{pages}{9360--9372}.
\newblock


\bibitem[\protect\citeauthoryear{Lu, Zhang, and Ma}{Lu et~al\mbox{.}}{2018}]%
        {Lu2018Between}
\bibfield{author}{\bibinfo{person}{Hongyu Lu}, \bibinfo{person}{Min Zhang},
  {and} \bibinfo{person}{Shaoping Ma}.} \bibinfo{year}{2018}\natexlab{}.
\newblock \showarticletitle{Between Clicks and Satisfaction: Study on
  Multi-Phase User Preferences and Satisfaction for Online News Reading}. In
  \bibinfo{booktitle}{{\em SIGIR}}. \bibinfo{publisher}{ACM},
  \bibinfo{pages}{435--444}.
\newblock


\bibitem[\protect\citeauthoryear{Lu, Zhang, Ma, Wang, xia, Liu, Lin, and Ma}{Lu
  et~al\mbox{.}}{2019}]%
        {Lu2019effects}
\bibfield{author}{\bibinfo{person}{Hongyu Lu}, \bibinfo{person}{Min Zhang},
  \bibinfo{person}{Weizhi Ma}, \bibinfo{person}{Ce Wang}, \bibinfo{person}{Feng
  xia}, \bibinfo{person}{Yiqun Liu}, \bibinfo{person}{Leyu Lin}, {and}
  \bibinfo{person}{Shaoping Ma}.} \bibinfo{year}{2019}\natexlab{}.
\newblock \showarticletitle{Effects of User Negative Experience in Mobile News
  Streaming}. In \bibinfo{booktitle}{{\em SIGIR}}. \bibinfo{publisher}{ACM},
  \bibinfo{pages}{705--714}.
\newblock


\bibitem[\protect\citeauthoryear{Morik, Singh, Hong, and Joachims}{Morik
  et~al\mbox{.}}{2020}]%
        {Marco2020Controlling}
\bibfield{author}{\bibinfo{person}{Marco Morik}, \bibinfo{person}{Ashudeep
  Singh}, \bibinfo{person}{Jessica Hong}, {and} \bibinfo{person}{Thorsten
  Joachims}.} \bibinfo{year}{2020}\natexlab{}.
\newblock \showarticletitle{Controlling Fairness and Bias in Dynamic
  Learning-to-Rank}. In \bibinfo{booktitle}{{\em SIGIR}}.
  \bibinfo{publisher}{ACM}, \bibinfo{pages}{429--438}.
\newblock


\bibitem[\protect\citeauthoryear{Niu, Tang, Zhang, Lu, Hua, and Wen}{Niu
  et~al\mbox{.}}{2020}]%
        {niu2020counterfactual}
\bibfield{author}{\bibinfo{person}{Yulei Niu}, \bibinfo{person}{Kaihua Tang},
  \bibinfo{person}{Hanwang Zhang}, \bibinfo{person}{Zhiwu Lu},
  \bibinfo{person}{Xian-Sheng Hua}, {and} \bibinfo{person}{Ji-Rong Wen}.}
  \bibinfo{year}{2020}\natexlab{}.
\newblock \showarticletitle{Counterfactual VQA: A Cause-Effect Look at Language
  Bias}. In \bibinfo{booktitle}{{\em arXiv:2006.04315}}.
\newblock


\bibitem[\protect\citeauthoryear{Ovaisi, Ahsan, Zhang, Vasilaky, and
  Zheleva}{Ovaisi et~al\mbox{.}}{2020}]%
        {Ovaisi2020Correcting}
\bibfield{author}{\bibinfo{person}{Zohreh Ovaisi}, \bibinfo{person}{Ragib
  Ahsan}, \bibinfo{person}{Yifan Zhang}, \bibinfo{person}{Kathryn Vasilaky},
  {and} \bibinfo{person}{Elena Zheleva}.} \bibinfo{year}{2020}\natexlab{}.
\newblock \showarticletitle{Correcting for Selection Bias in Learning-to-Rank
  Systems}. In \bibinfo{booktitle}{{\em WWW}}. \bibinfo{publisher}{ACM},
  \bibinfo{pages}{1863--1873}.
\newblock


\bibitem[\protect\citeauthoryear{Pearl}{Pearl}{2001}]%
        {judea2001direct}
\bibfield{author}{\bibinfo{person}{Judea Pearl}.}
  \bibinfo{year}{2001}\natexlab{}.
\newblock \showarticletitle{Direct and indirect effects}. In
  \bibinfo{booktitle}{{\em UAI}}. \bibinfo{publisher}{Morgan Kaufmann
  Publishers Inc}, \bibinfo{pages}{411--420}.
\newblock


\bibitem[\protect\citeauthoryear{Pearl}{Pearl}{2009}]%
        {pearl2009causality}
\bibfield{author}{\bibinfo{person}{Judea Pearl}.}
  \bibinfo{year}{2009}\natexlab{}.
\newblock \bibinfo{booktitle}{{\em Causality}}.
\newblock \bibinfo{publisher}{Cambridge university press}.
\newblock


\bibitem[\protect\citeauthoryear{Pearl and Mackenzie}{Pearl and
  Mackenzie}{2018}]%
        {Pearl2018the}
\bibfield{author}{\bibinfo{person}{Judea Pearl} {and} \bibinfo{person}{Dana
  Mackenzie}.} \bibinfo{year}{2018}\natexlab{}.
\newblock \bibinfo{booktitle}{{\em The Book of Why: The New Science of Cause
  and Effect\/} (\bibinfo{edition}{1st} ed.)}.
\newblock \bibinfo{publisher}{Basic Books, Inc.}
\newblock


\bibitem[\protect\citeauthoryear{Rendle, Freudenthaler, Gantner, and
  Schmidt-Thieme}{Rendle et~al\mbox{.}}{2009}]%
        {rendle2009bpr}
\bibfield{author}{\bibinfo{person}{Steffen Rendle}, \bibinfo{person}{Christoph
  Freudenthaler}, \bibinfo{person}{Zeno Gantner}, {and} \bibinfo{person}{Lars
  Schmidt-Thieme}.} \bibinfo{year}{2009}\natexlab{}.
\newblock \showarticletitle{BPR: Bayesian personalized ranking from implicit
  feedback}. In \bibinfo{booktitle}{{\em UAI}}. AUAI Press,
  \bibinfo{pages}{452--461}.
\newblock


\bibitem[\protect\citeauthoryear{Rosenbaum and Rubin}{Rosenbaum and
  Rubin}{1983}]%
        {ROSENBAUM1983the}
\bibfield{author}{\bibinfo{person}{Paul~R. Rosenbaum} {and}
  \bibinfo{person}{Donald~B. Rubin}.} \bibinfo{year}{1983}\natexlab{}.
\newblock \showarticletitle{{The central role of the propensity score in
  observational studies for causal effects}}.
\newblock \bibinfo{journal}{{\em Biometrika\/}} \bibinfo{volume}{70},
  \bibinfo{number}{1} (\bibinfo{date}{04} \bibinfo{year}{1983}),
  \bibinfo{pages}{41--55}.
\newblock


\bibitem[\protect\citeauthoryear{Saito, Yaginuma, Nishino, Sakata, and
  Nakata}{Saito et~al\mbox{.}}{2020}]%
        {saito2020unbiased}
\bibfield{author}{\bibinfo{person}{Yuta Saito}, \bibinfo{person}{Suguru
  Yaginuma}, \bibinfo{person}{Yuta Nishino}, \bibinfo{person}{Hayato Sakata},
  {and} \bibinfo{person}{Kazuhide Nakata}.} \bibinfo{year}{2020}\natexlab{}.
\newblock \showarticletitle{Unbiased Recommender Learning from
  Missing-Not-At-Random Implicit Feedback}. In \bibinfo{booktitle}{{\em WSDM}}.
  \bibinfo{publisher}{ACM}, \bibinfo{pages}{501--509}.
\newblock


\bibitem[\protect\citeauthoryear{Tang, Huang, and Zhang}{Tang
  et~al\mbox{.}}{2020a}]%
        {tang2020longtailed}
\bibfield{author}{\bibinfo{person}{Kaihua Tang}, \bibinfo{person}{Jianqiang
  Huang}, {and} \bibinfo{person}{Hanwang Zhang}.}
  \bibinfo{year}{2020}\natexlab{a}.
\newblock \showarticletitle{Long-Tailed Classification by Keeping the Good and
  Removing the Bad Momentum Causal Effect}. In \bibinfo{booktitle}{{\em
  NeurIPS}}.
\newblock


\bibitem[\protect\citeauthoryear{Tang, Niu, Huang, Shi, and Zhang}{Tang
  et~al\mbox{.}}{2020b}]%
        {tang2020unbiased}
\bibfield{author}{\bibinfo{person}{Kaihua Tang}, \bibinfo{person}{Yulei Niu},
  \bibinfo{person}{Jianqiang Huang}, \bibinfo{person}{Jiaxin Shi}, {and}
  \bibinfo{person}{Hanwang Zhang}.} \bibinfo{year}{2020}\natexlab{b}.
\newblock \showarticletitle{Unbiased scene graph generation from biased
  training}. In \bibinfo{booktitle}{{\em arXiv:2002.11949}}.
\newblock


\bibitem[\protect\citeauthoryear{VanderWeele}{VanderWeele}{2013}]%
        {vanderweele2013three}
\bibfield{author}{\bibinfo{person}{Tyler~J VanderWeele}.}
  \bibinfo{year}{2013}\natexlab{}.
\newblock \showarticletitle{A three-way decomposition of a total effect into
  direct, indirect, and interactive effects}.
\newblock \bibinfo{journal}{{\em Epidemiology (Cambridge, Mass.)\/}}
  \bibinfo{volume}{24}, \bibinfo{number}{2} (\bibinfo{year}{2013}),
  \bibinfo{pages}{224}.
\newblock


\bibitem[\protect\citeauthoryear{Wang, Duan, Jiang, Jing, Song, and Nie}{Wang
  et~al\mbox{.}}{2021a}]%
        {Wang2021Market2Dish}
\bibfield{author}{\bibinfo{person}{Wenjie Wang}, \bibinfo{person}{Ling-Yu
  Duan}, \bibinfo{person}{Hao Jiang}, \bibinfo{person}{Peiguang Jing},
  \bibinfo{person}{Xuemeng Song}, {and} \bibinfo{person}{Liqiang Nie}.}
  \bibinfo{year}{2021}\natexlab{a}.
\newblock \showarticletitle{Market2Dish: Health-Aware Food Recommendation}.
\newblock \bibinfo{journal}{{\em TOMM\/}}  \bibinfo{volume}{17}
  (\bibinfo{date}{April} \bibinfo{year}{2021}).
\newblock


\bibitem[\protect\citeauthoryear{Wang, Feng, He, Nie, and Chua}{Wang
  et~al\mbox{.}}{2021b}]%
        {wang2021denoising}
\bibfield{author}{\bibinfo{person}{Wenjie Wang}, \bibinfo{person}{Fuli Feng},
  \bibinfo{person}{Xiangnan He}, \bibinfo{person}{Liqiang Nie}, {and}
  \bibinfo{person}{Tat-Seng Chua}.} \bibinfo{year}{2021}\natexlab{b}.
\newblock \showarticletitle{Denoising implicit feedback for recommendation}. In
  \bibinfo{booktitle}{{\em WSDM}}. \bibinfo{publisher}{ACM},
  \bibinfo{pages}{373--381}.
\newblock


\bibitem[\protect\citeauthoryear{Wang, He, Wang, Feng, and Chua}{Wang
  et~al\mbox{.}}{2019}]%
        {wang2019NGCF}
\bibfield{author}{\bibinfo{person}{Xiang Wang}, \bibinfo{person}{Xiangnan He},
  \bibinfo{person}{Meng Wang}, \bibinfo{person}{Fuli Feng}, {and}
  \bibinfo{person}{Tat{-}Seng Chua}.} \bibinfo{year}{2019}\natexlab{}.
\newblock \showarticletitle{Neural Graph Collaborative Filtering}. In
  \bibinfo{booktitle}{{\em SIGIR}}. \bibinfo{publisher}{ACM},
  \bibinfo{pages}{165--174}.
\newblock


\bibitem[\protect\citeauthoryear{Wei, Wang, Nie, He, and Chua}{Wei
  et~al\mbox{.}}{2020}]%
        {wei2020graph}
\bibfield{author}{\bibinfo{person}{Yinwei Wei}, \bibinfo{person}{Xiang Wang},
  \bibinfo{person}{Liqiang Nie}, \bibinfo{person}{Xiangnan He}, {and}
  \bibinfo{person}{Tat-Seng Chua}.} \bibinfo{year}{2020}\natexlab{}.
\newblock \showarticletitle{Graph-Refined Convolutional Network for Multimedia
  Recommendation with Implicit Feedback}. In \bibinfo{booktitle}{{\em MM}}.
  \bibinfo{pages}{3541--3549}.
\newblock


\bibitem[\protect\citeauthoryear{Wei, Wang, Nie, He, Hong, and Chua}{Wei
  et~al\mbox{.}}{2019}]%
        {wei2019mmgcn}
\bibfield{author}{\bibinfo{person}{Yinwei Wei}, \bibinfo{person}{Xiang Wang},
  \bibinfo{person}{Liqiang Nie}, \bibinfo{person}{Xiangnan He},
  \bibinfo{person}{Richang Hong}, {and} \bibinfo{person}{Tat-Seng Chua}.}
  \bibinfo{year}{2019}\natexlab{}.
\newblock \showarticletitle{\text{MMGCN:} Multi-modal Graph Convolution Network
  for Personalized Recommendation of Micro-video}. In \bibinfo{booktitle}{{\em
  MM}}. \bibinfo{publisher}{ACM}, \bibinfo{pages}{1437--1445}.
\newblock


\bibitem[\protect\citeauthoryear{Wen, Yang, and Estrin}{Wen
  et~al\mbox{.}}{2019}]%
        {Wen2019Leveraging}
\bibfield{author}{\bibinfo{person}{Hongyi Wen}, \bibinfo{person}{Longqi Yang},
  {and} \bibinfo{person}{Deborah Estrin}.} \bibinfo{year}{2019}\natexlab{}.
\newblock \showarticletitle{Leveraging Post-click Feedback for Content
  Recommendations}. In \bibinfo{booktitle}{{\em RecSys}}.
  \bibinfo{publisher}{ACM}, \bibinfo{pages}{278--286}.
\newblock


\bibitem[\protect\citeauthoryear{Wen, Zhang, Wang, Lv, Bao, Lin, and Yang}{Wen
  et~al\mbox{.}}{2020}]%
        {wen2020entire}
\bibfield{author}{\bibinfo{person}{Hong Wen}, \bibinfo{person}{Jing Zhang},
  \bibinfo{person}{Yuan Wang}, \bibinfo{person}{Fuyu Lv},
  \bibinfo{person}{Wentian Bao}, \bibinfo{person}{Quan Lin}, {and}
  \bibinfo{person}{Keping Yang}.} \bibinfo{year}{2020}\natexlab{}.
\newblock \showarticletitle{Entire space multi-task modeling via post-click
  behavior decomposition for conversion rate prediction}. In
  \bibinfo{booktitle}{{\em SIGIR}}. \bibinfo{publisher}{ACM},
  \bibinfo{pages}{2377--2386}.
\newblock


\bibitem[\protect\citeauthoryear{Yang, Lee, Park, and goo Lee}{Yang
  et~al\mbox{.}}{2012}]%
        {Yang2012Exploiting}
\bibfield{author}{\bibinfo{person}{Byoungju Yang}, \bibinfo{person}{Sangkeun
  Lee}, \bibinfo{person}{Sungchan Park}, {and} \bibinfo{person}{Sang goo Lee}.}
  \bibinfo{year}{2012}\natexlab{}.
\newblock \showarticletitle{Exploiting Various Implicit Feedback for
  Collaborative Filtering}. In \bibinfo{booktitle}{{\em WWW}}.
  \bibinfo{publisher}{ACM}, \bibinfo{pages}{639--640}.
\newblock


\bibitem[\protect\citeauthoryear{Yi, Hong, Zhong, Liu, and Rajan}{Yi
  et~al\mbox{.}}{2014}]%
        {Yi2014Beyond}
\bibfield{author}{\bibinfo{person}{Xing Yi}, \bibinfo{person}{Liangjie Hong},
  \bibinfo{person}{Erheng Zhong}, \bibinfo{person}{Nanthan~Nan Liu}, {and}
  \bibinfo{person}{Suju Rajan}.} \bibinfo{year}{2014}\natexlab{}.
\newblock \showarticletitle{Beyond clicks: dwell time for personalization}. In
  \bibinfo{booktitle}{{\em RecSys}}. \bibinfo{publisher}{ACM},
  \bibinfo{pages}{113--120}.
\newblock


\bibitem[\protect\citeauthoryear{Yin, Luo, Lee, and Wang}{Yin
  et~al\mbox{.}}{2013}]%
        {Yin2013silence}
\bibfield{author}{\bibinfo{person}{Peifeng Yin}, \bibinfo{person}{Ping Luo},
  \bibinfo{person}{Wang-Chien Lee}, {and} \bibinfo{person}{Min Wang}.}
  \bibinfo{year}{2013}\natexlab{}.
\newblock \showarticletitle{Silence is Also Evidence: Interpreting Dwell Time
  for Recommendation from Psychological Perspective}. In
  \bibinfo{booktitle}{{\em KDD}}. \bibinfo{publisher}{ACM},
  \bibinfo{pages}{989--997}.
\newblock


\bibitem[\protect\citeauthoryear{Yuan, He, Karatzoglou, and Zhang}{Yuan
  et~al\mbox{.}}{2020}]%
        {yuan2020parameter}
\bibfield{author}{\bibinfo{person}{Fajie Yuan}, \bibinfo{person}{Xiangnan He},
  \bibinfo{person}{Alexandros Karatzoglou}, {and} \bibinfo{person}{Liguang
  Zhang}.} \bibinfo{year}{2020}\natexlab{}.
\newblock \showarticletitle{Parameter-efficient transfer from sequential
  behaviors for user modeling and recommendation}. In \bibinfo{booktitle}{{\em
  SIGIR}}. \bibinfo{publisher}{ACM}, \bibinfo{pages}{1469--1478}.
\newblock


\bibitem[\protect\citeauthoryear{Yue, Patel, and Roehrig}{Yue
  et~al\mbox{.}}{2010}]%
        {Yue2010Beyond}
\bibfield{author}{\bibinfo{person}{Yisong Yue}, \bibinfo{person}{Rajan Patel},
  {and} \bibinfo{person}{Hein Roehrig}.} \bibinfo{year}{2010}\natexlab{}.
\newblock \showarticletitle{Beyond Position Bias: Examining Result
  Attractiveness as a Source of Presentation Bias in Clickthrough Data}. In
  \bibinfo{booktitle}{{\em WWW}}. \bibinfo{publisher}{ACM},
  \bibinfo{pages}{1011--1018}.
\newblock


\bibitem[\protect\citeauthoryear{Zannettou, Chatzis, Papadamou, and
  Sirivianos}{Zannettou et~al\mbox{.}}{2018}]%
        {zannettou2018good}
\bibfield{author}{\bibinfo{person}{Savvas Zannettou}, \bibinfo{person}{Sotirios
  Chatzis}, \bibinfo{person}{Kostantinos Papadamou}, {and}
  \bibinfo{person}{Michael Sirivianos}.} \bibinfo{year}{2018}\natexlab{}.
\newblock \showarticletitle{The good, the bad and the bait: Detecting and
  characterizing clickbait on YouTube}. In \bibinfo{booktitle}{{\em SPW}}.
  IEEE, \bibinfo{pages}{63--69}.
\newblock


\bibitem[\protect\citeauthoryear{Zhang, Bao, Liu, Yang, Lin, Wen, and
  Ramezani}{Zhang et~al\mbox{.}}{2020}]%
        {Zhang2020Large}
\bibfield{author}{\bibinfo{person}{Wenhao Zhang}, \bibinfo{person}{Wentian
  Bao}, \bibinfo{person}{Xiao-Yang Liu}, \bibinfo{person}{Keping Yang},
  \bibinfo{person}{Quan Lin}, \bibinfo{person}{Hong Wen}, {and}
  \bibinfo{person}{Ramin Ramezani}.} \bibinfo{year}{2020}\natexlab{}.
\newblock \showarticletitle{Large-Scale Causal Approaches to Debiasing
  Post-Click Conversion Rate Estimation with Multi-Task Learning}. In
  \bibinfo{booktitle}{{\em WWW}}. \bibinfo{publisher}{ACM},
  \bibinfo{pages}{2775--2781}.
\newblock


\bibitem[\protect\citeauthoryear{Zhu, He, Zhang, and Caverlee}{Zhu
  et~al\mbox{.}}{2020}]%
        {Zhu2020unbiased}
\bibfield{author}{\bibinfo{person}{Ziwei Zhu}, \bibinfo{person}{Yun He},
  \bibinfo{person}{Yin Zhang}, {and} \bibinfo{person}{James Caverlee}.}
  \bibinfo{year}{2020}\natexlab{}.
\newblock \showarticletitle{Unbiased Implicit Recommendation and Propensity
  Estimation via Combinational Joint Learning}. In \bibinfo{booktitle}{{\em
  RecSys}}. \bibinfo{publisher}{ACM}, \bibinfo{pages}{551--556}.
\newblock


\end{thebibliography}
}

\end{document}